\documentclass[letterpaper,twocolumn,10pt]{article}
\usepackage{usenix-2020-09}
\usepackage{tikz}
\usepackage{amsmath}
\usepackage{comment}
\usepackage{booktabs} 
\usepackage{xspace}
\usepackage{graphicx}
\usepackage{subfigure}
\usepackage{makecell}
\usepackage{amssymb}
\usepackage{diagbox}
\usepackage{listings}
\usepackage{graphics}
\usepackage{epsfig}
\usepackage{color}
\usepackage{ifpdf}
\usepackage{mathrsfs}
\usepackage{multirow}
\usepackage{xcolor}
\usepackage{url}
\usepackage{color,xcolor}
\usepackage{array}
\usepackage{multirow}
\usepackage{bm}
\usepackage{subfigure}

\begin{document}

\date{}

\title{\Large \bf Your Microphone Array Retains Your Identity:\\ A Robust Voice Liveness Detection System for Smart Speakers}

\author{
{\rm Yan Meng\textsuperscript{1}, Jiachun Li\textsuperscript{1}, Matthew Pillari\textsuperscript{2}, Arjun Deopujari\textsuperscript{2}} \\
{\rm Liam Brennan\textsuperscript{2}, Hafsah Shamsie\textsuperscript{2}, Haojin Zhu\textsuperscript{1}, and Yuan Tian\textsuperscript{2}}\\
{\textsuperscript{1}Shanghai Jiao Tong University, \{yan\_meng, jiachunli, zhu-hj\}@sjtu.edu.cn}\\
{\textsuperscript{2}University of Virginia, \{pillari, ajd4mq, lfb6ek, his3uh, yuant\}@virginia.edu}\\
}

\maketitle

\def\UrlBreaks{\do\A\do\B\do\C\do\D\do\E\do\F\do\G\do\H\do\I\do\J
\do\K\do\L\do\M\do\N\do\O\do\P\do\Q\do\R\do\S\do\T\do\U\do\V
\do\W\do\X\do\Y\do\Z\do\[\do\\\do\]\do\^\do\_\do\`\do\a\do\b
\do\c\do\d\do\e\do\f\do\g\do\h\do\i\do\j\do\k\do\l\do\m\do\n
\do\o\do\p\do\q\do\r\do\s\do\t\do\u\do\v\do\w\do\x\do\y\do\z
\do\.\do\@\do\\\do\/\do\!\do\_\do\|\do\;\do\>\do\]\do\)\do\,
\do\?\do\'\do+\do\=\do\#}

\newcommand{\frameName}{\textsc{ArrayID}\xspace}
\newcommand{\void}{\textsc{Void}\xspace}
\newcommand{\cafield}{\textsc{CaField}\xspace}
\newcommand{\print}{\textsc{Array-print}\xspace}

\newcommand{\eg}{\emph{e.g.}\xspace}
\newcommand{\etal}{\emph{et al.}\xspace}
\newcommand{\etc}{\emph{etc}\xspace}
\newcommand{\ie}{\emph{i.e.}\xspace}

\begin{abstract}
Though playing an essential role in smart home systems, smart speakers are vulnerable to voice spoofing attacks. Passive liveness detection, which utilizes only the collected audio rather than the deployed sensors to distinguish between live-human and replayed voices, has drawn increasing attention.
However, it faces the challenge of performance degradation under the different environmental factors as well as the strict requirement of the fixed user gestures. 

In this study, we propose a novel liveness feature, \textit{array fingerprint}, which utilizes the microphone array inherently adopted by the smart speaker to determine the identity of collected audios. Our theoretical analysis demonstrates that by leveraging the circular layout of microphones, compared with existing schemes, array fingerprint achieves a more robust performance under the environmental change and user's movement. Then, to leverage such a fingerprint, we propose \frameName, a lightweight passive detection scheme, and elaborate a series of features working together with array fingerprint. 
Our evaluation on the dataset containing 32,780 audio samples and 14 spoofing devices shows that \frameName achieves an accuracy of 99.84\%, which is superior to existing passive liveness detection schemes. 
\end{abstract}

\pagenumbering{gobble}

\section{Introduction}

Nowadays, voice assistance-enabled smart speakers serve as the hub of popular smart home platforms (\eg, Amazon Alexa, Google Home) and allow the user to remotely control home appliances (\eg, smart lighter, locker, thermostat) or query information (\eg, weather, news) as long as it can hear the user. 
However, the inherent broadcast nature of voice unlocks a door for adversaries to inject malicious commands (\ie, spoofing attack). 
Besides the classical replay attack \cite{Diao:2014:YVA:2666620.2666623, wangccs2020},  emerging attacks leveraging flaws in smart speakers are also proposed by researchers. On the hardware side, the non-linearity of the microphone's frequency response provides a door for inaudible ultrasound-based attacks (\eg, \textit{Dolphin attack} \cite{dolphinAttack} and \textit{BackDoor attack} \cite{BackDoor}). For the software aspect, the deep learning models employed by both speech recognition and speaker verification are proved to be vulnerable to emerging adversarial attacks such as hidden voice~\cite{2016Nicholas}, CommanderSong~\cite{commandsong}, and user impersonation~\cite{vmask}.
Spoofing attacks impose emerging safety issues (\eg, deliberately turn on the smart thermostat~\cite{DBLP:conf/ccs/DingH18}) and privacy risks (\eg, querying user's schedule information) on the smart speaker and therefore cause great concern. 

To defend against spoofing attacks, researchers have proposed a variety of countermeasures. Almost all countermeasures leverage the fact that voices in the spoofing attack are played by electrical devices (\eg, high-quality loudspeaker~\cite{wangccs2020}, ultrasonic dynamic speaker~\cite{dolphinAttack}). Thus, the physical characteristics, which are different between humans and machines, could be used as the ``liveness'' factors. The existing countermeasures (\textit{aka.}, liveness detection) could be divided into 
multi-factor authentication and passive scheme. The former combines the collected audio and additional physical quantity (\eg, acceleration~\cite{vauth}, electromagnetic field~\cite{magnetic}, ultrasound~\cite{Zhang:2017:HYV:3133956.3133962}, Wi-Fi~\cite{Meng:2018:WES:3209582.3209591}, mm-Wave~\cite{9193961}) to distinguish between the human voice and the machine-generated one. 
By contrast, the passive scheme only considers the audio data collected by the smart speaker. Its key insight is that the difference of articulatory manners between real humans (\ie, vocal vibration and mouth movement) and electrical machines (\ie, diaphragm vibration) will result in the subtle but significant differences in the collected audios' spectrograms. 
Passive schemes based on mono audio~\cite{blue2018hello, void2020} and two-channel audio~\cite{blue20182ma, fieldprint2019} have already been proposed and could be directly incorporated in the smart speaker's software level.

However, the existing liveness detection schemes face a series of challenges in the aspects of \textit{usability} and \textit{efficiency}, which seriously hinder their real-world deployment in practice. 
On the one hand, to capture the liveness factor of a real human, multiple-factor authentication either requires the user to carry specialized sensors (\eg, accelerator, magnetometer) or actively emits probe signals (\eg, ultrasounds, wireless signals), which adds additional burdens for users. 
On the other hand, passive schemes leveraging sub-bass low-frequency area (20\textasciitilde 300 Hz in \cite{blue2018hello}) or voice area (below 10 kHz in ~\cite{void2020}) of mono audio's spectrum as liveness factor are vulnerable to sound propagation channel's change and the spectrum modulated-based attack~\cite{wangccs2020}. Another scheme \cite{fieldprint2019} aiming to extract audio's \textit{fieldprint}
from two-channel audio requires the user to keep a fixed manner to ensure the robustness of such fingerprints. As a result, the scheme is difficult to be deployed in many practical scenarios (\eg, users walking or having gesture changes). 
Therefore, it is desirable to propose a novel passive liveness detection scheme with the following merits:  \textit{(\romannumeral1) Device-free:} performing passive detection only relying on the collected audio;  \textit{(\romannumeral2) Resilient to environment change:} being robust to dynamic sound propagation channel and user's movement,  \textit{(\romannumeral3) High accuracy:} achieving high accuracy compared to existing works.  

\textbf{Motivations.} To achieve a device-free, robust passive liveness detection, in this study, we propose \frameName, a microphone array-based liveness detection system, to effectively defend against spoofing attacks. \frameName is motivated from the basic observation that the microphone array has been widely adopted by the mainstream smart speakers (\eg, both of Amazon Echo 3rd Gen~\cite{alexa:review} and Google Home Max~\cite{google:review} having 6 microphones), which is expected to significantly enhance the diversity of the collected audios thanks to the different locations and mutual distances of the microphones in this array. By exploiting the audio diversity, \frameName can extract more useful information related to the target user, which is expected to significantly improve the robustness and accuracy of the liveness detection.

\textbf{Challenges.} To implement this basic idea, this study tries to address the following three key challenges: \textit{(\romannumeral1)} Theoretically, what is the advantage of adopting a microphone array compared with a single microphone? \textit{(\romannumeral2)} Considering the dynamic audio propagation channel, how can we eliminate the distortions caused by environment factors (\eg, dynamic air channel and user's position changes) by leveraging the microphone array? \textit{(\romannumeral3)} Considering that our work is the first one to leverage microphone array for liveness detection and there is no large-scale microphone array-based indoor audio dataset available so far, how can we demonstrate the effectiveness and accuracy of the proposed scheme?

To solve the above three problems, we first build a sound propagation model based on the wave propagation theory and then leverage it to theoretically assess the impact of environment factors (\eg, articulatory gesture, sound decay pattern, propagation path) on the final collected audio's spectrum. Secondly, after collecting multi-channel audio, we give a formal definition of array fingerprint and discuss the theoretic performance gain of adopting microphone array, which can leverage the relationship among different channels' data to eliminate the distortions caused by factors including air channel and user's position changes. 
Thirdly, we collect and build the first array fingerprint-based open dataset containing multi-channel voices from 38,720 voice commands. 
To evaluate the effectiveness of \frameName, we compare \frameName with previous passive schemes (\ie, \cafield~\cite{fieldprint2019}, and \void~\cite{void2020}) on both our dataset and a third-party dataset called ReMasc Core dataset~\cite{remasc}.
\frameName achieves the authentication accuracy of 99.84\% and 97.78\%  on our dataset and ReMasc Core dataset, respectively, while the best performance of existing schemes~\cite{void2020, fieldprint2019} on these two datasets are 98.81\% and 84.37\% respectively. The experimental results well demonstrate the effectiveness and robustness of \frameName.

To the best of our knowledge, our work is the first to exploit the circular microphone array of the smart speaker to perform passive liveness detection in a smart home environment. The contributions of this study are summarized as follows:
\begin{itemize}
    \item \textit{Novel system.} We design, implement and evaluate \frameName for thwarting voice spoofing attacks. By only using audio collected from a smart speaker, \frameName does not require the user to carry any device or conduct additional action.
    \item \textit{Effective feature.} We give a theoretical analysis of principles behind passive detection and propose a robust liveness feature: the array fingerprint. This novel feature both enhances effectiveness and broadens the application scenarios of passive liveness detection.
    \item \textit{Robust performance.} 
    Experimental results on both our dataset and a third-party dataset show that \frameName outperforms existing schemes. Besides, we evaluate multiple factors (\eg, distance, direction, spoofing devices, noise) and demonstrate the robustness of \frameName.
    \item \textit{New large-scale dataset.} A dataset containing 14 different spoofing devices collected by microphone array will be available to researchers, vendors, and developers for evaluating further liveness detection schemes. 

\end{itemize}

The rest of this paper is organized as follows. In Section~\ref{sec:preliminaries}, we introduce the preliminaries of this study. In Section~\ref{sec:motivation}, we propose the concept of the array fingerprint and proof its advantages by both theoretical analysis and experiments. We elaborate on the detailed design of \frameName in Section~\ref{sec:design}, which is followed by evaluation, discussion, and related work in Sections~\ref{sec:evaluation},~\ref{sec:discussions}, and~\ref{sec:related:work}, respectively. Finally, we conclude this paper in Section~\ref{sec:conclusion}.

\section{Preliminaries}
\label{sec:preliminaries}

\begin{figure*}[ht]
\centering
\subfigure[Sound generation.]{
\includegraphics[height = 2cm]{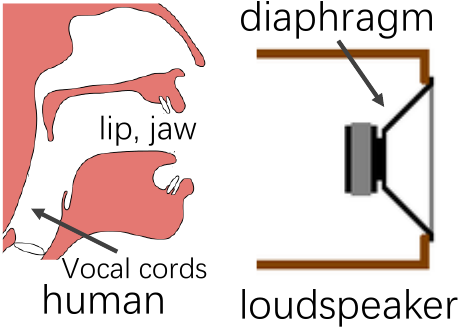}
\label{fig:sound:generation}
}
\subfigure[Sound propagation process.]{
\includegraphics[height = 2cm]{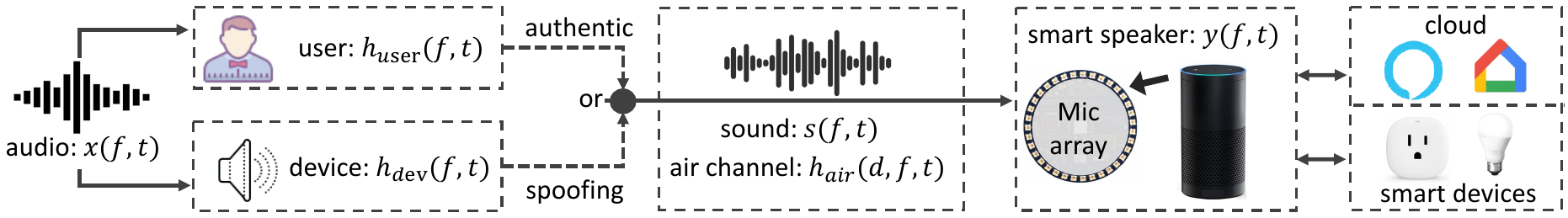}
\label{fig:sound:propagation}
}
\caption{Sound generation and propagation in smart home.}
\vspace{-4mm}
\label{fig:sound}
\end{figure*}

\subsection{Threat Model}

In this study, we focus on the voice spoofing attack, which can be categorized into the following two types.

\textbf{Classical replay attacks.}
To fool the voice assistance, the attacker collects the legitimate user's audio samples and then plays it back with a high-quality loudspeaker \cite{Diao:2014:YVA:2666620.2666623}. The victim's voice audio can be recorded or captured in many manners, which is not limited to websites, daily life talking, phone calls, etc. The replay attack is the most effective among various spoofing approaches since it preserves the most comprehensive voiceprint of the victim and requires no cumbersome hardware configurations and software parameter fine-tuning.

\textbf{Advanced adversarial attacks.} Even if attackers can only collect a limited number of the target user's voice samples, by adopting the latest voice synthesized technique~\cite{who:is:real:bob}, it is still feasible to attack existing speech recognition and speaker verification systems. For instance, the adversary can craft subtle noises into the audio (\eg, hidden voice ~\cite{2016Nicholas}, music ~\cite{commandsong} or a broadcaster's voice~\cite{vmask}) or inaudible ultrasounds~\cite{dolphinAttack, BackDoor} to launch an attack without raising the victim's concern. Moreover, by carefully modifying the spectrum of spoofing audio, the modulated attack~\cite{wangccs2020} proposed by Wang \etal, demonstrates the feasibility of bypassing existing mono audio-based liveness detection schemes~\cite{blue2018hello}.

Similar to the previous works~\cite{blue2018hello, fieldprint2019, void2020}, in this study, the adversary is assumed to already obtain the victim's audio samples and can remotely control the victim's audio device (\eg, smart TV, smartphone) to launch the voice spoofing attack.
In this study, we mainly investigate how to leverage passive liveness detection to thwart replay attacks since most of the existing voice biometric-based
authentication (human speaker verification) systems are vulnerable to this kind of replay attack. We also study \frameName's performance on thwarting advanced attacks including modulated attack~\cite{wangccs2020},
hidden voice~\cite{2016Nicholas}, and VMask~\cite{vmask} in Section~\ref{sec:eval:advanced}.

\subsection{Sound Generation and Propagation}
\label{sec:sound:generation}
Before reviewing existing passive liveness detection schemes, it is important to describe the sound generation and propagation process. 

\textbf{Sound generation.} 
As shown in Figure~\ref{fig:sound:generation}, voice commands are generated by a human or electrical device (\ie, loudspeaker). For the loudspeaker, given an original voice command signal $x(f,t)$, where $f$ represents the frequency and $t$ is time, the loudspeaker utilizes the electromagnetic field change to vibrate the diaphragm. The movement of the diaphragm suspends and pushes air to generate the sound wave $s(f,t) = h_{dev}(f,t)\cdot x(f,t)$, where $h_{dev}(f,t)$ represents the channel gain in the sound signal modulation by the device as shown in Figure~\ref{fig:sound:propagation}. Similarly, when a user speaks voice commands, their mouth and lips also modulate the air and we can use $h_{user}(f,t)$ to represent the modulation gain, where the generated sound is $s(f,t) = h_{user}(f,t)\cdot x(f,t)$ \footnote{In the real-world scenario, there is no such $x(f,t)$ during human voice generation process. However, the concepts of $x(f,t)$ and $h_{user}(f,t)$ are widely used~\cite{blue2018hello} and will help us understand features in Section \ref{sec:feature}.}. 
Finally, the generated sound $s(f,t)$ is spread through the air and captured by the smart speaker. 

\textbf{Sound transmission.}
Currently, smart speakers usually have a microphone array (\eg, Amazon Echo 3rd Gen~\cite{alexa:review} and Google Home Max~\cite{google:review} both have 6 microphones). For a given microphone, when sound is transmitted to it, the air pressure at the microphone's location can be represented as $y(f,t) = h_{air}(d,f,t)\cdot s(f,t)$, where $d$ is the distance of the transmission path between the audio source and the microphone and $h_{air}(d,f,t)$ is the channel gain in the air propagation of the sound signal.

\textbf{Sound processing within the smart speaker.}
Finally, $y(f,t)$ is converted to an electrical signal by the microphone.
Since the microphones employed by mainstream smart speakers usually have a flat frequency response curve in the frequency area of the human voice, 
we assume smart speakers save original sensed data $y(f,t)$ which is also adopted by existing studies~\cite{fieldprint2019}. Finally, the collected audio signal is uploaded to the smart home cloud to further influence the actions of smart devices.

\subsection{Passive Liveness Detection}
\label{sec:passive:liveness}

The recently proposed liveness detection schemes could be divided into two categories: mono channel-based detection (\eg, Sub-bass~\cite{blue2018hello} and VOID~\cite{void2020}) and fieldprint-based detection (\ie, \cafield~\cite{fieldprint2019}).

\subsubsection{Mono Channel-based Detection}

\begin{figure}[t]
	\centering
	\subfigure[Spectrums on room A.]{
		\label{fft_location_a}
		\includegraphics[height=2.3cm]{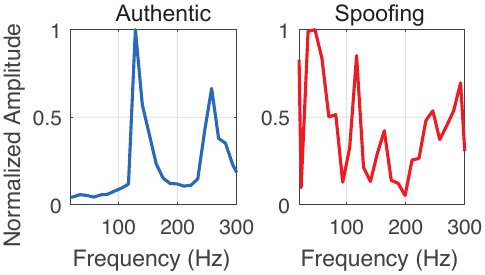}}
	\subfigure[Spectrums on room B.]{
		\label{fft_location_b}
		\includegraphics[height=2.3cm]{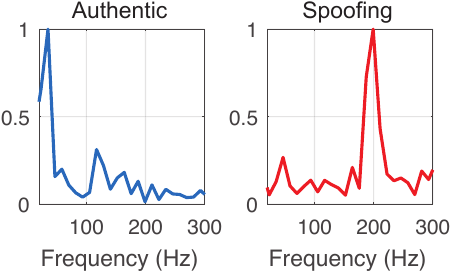}}
	\caption{Spectrums of authentic and spoofing voices when putting the smart speaker at different rooms.}
\vspace{-4mm}
	\label{level}
\end{figure}

\textbf{Principles}. As shown in Figure~\ref{fig:sound:generation}, the different sound generation principles between real human and electrical spoofing devices could be characterized as two different filters: $h_{user}(f,t)$  and $h_{dev}(f,t)$. If ignoring the distortion in the sound signal transmission, $h_{air}(d,f,t)$ could be considered as a constant value $A$. Thus, the received audio samples in authentic and spoofing attack scenarios are $y_{auth}(d,f,t) = A\cdot h_{user}(f,t)\cdot x(f,t)$ and $y_{spoof}(d,f,t) = A\cdot h_{dev}(f,t)\cdot x(f,t)$, respectively. Since $A$ and $x(f,t)$ are the same, it means that the spectrograms of the received audio samples already contain the identity of the audio source (the real user $h_{user}(f,t)$ or the spoofing one $h_{dev}(f,t)$). Figure~\ref{fft_location_a} shows the spectrums of the voice command ``OK Google'' and its spoofing counterpart. It's observed that the sub-bass spectrum (20-300 Hz) between two audio samples are quite different even if they are deemed similar, and this phenomenon is utilized by mono channel-based schemes such as Sub-base~\cite{blue2018hello}.

\noindent \textbf{Limitations.} However, in a real-world environment, $h_{air}(d, f, t)$ cannot always be regarded as a constant. The surrounding object's shape and materials, the sound transmission path, and the absorption coefficient of air all affect the value of $h_{air}(d, f, t)$. As shown in Figure~\ref{fft_location_a} and Figure~\ref{fft_location_b}, the spectrograms of authentic and spoof audio samples change drastically when putting the smart speaker in different rooms. 
The experimental result from Section~\ref{sec:eval:performance} and~\cite{void2020} demonstrates the performance of liveness detection undergoes degradation when handling datasets in which audios are collected from complicated environments (\eg, ASVSpoofing 2017 Challenge~\cite{Kinnunen2017}, ReMasc Core~\cite{remasc}).

\subsubsection{Fieldprint-based Detection}

\textbf{Principles}. 
The concept of \textit{fieldprint}~\cite{fieldprint2019} is based on the assumption that audio sources with different articulatory behaviors will cause a unique ``sound field'' around them. By measuring the field characteristics around the audio source, it is feasible to induce the audio's identity. \cafield is the typical scheme which deploys two microphones to receive two audios $y_1(f,t)$ and $y_2(f,t)$, and defines the fieldprint as:
\begin{equation}
    Field = log(\frac{y_1(f,t)}{y_2(f,t)}).
    \label{eq:field-print}
\end{equation}

\noindent \textbf{Limitations.} Measuring stable and accurate fieldprint requires the position between the audio source and the print measure sensors must be relatively stable. For instance, \cafield only performs well when the user holds a smartphone equipped with two microphones close to the face in a fixed manner. The fieldprint struggles in far distances (\eg, greater than 40 cm in~\cite{fieldprint2019}), making it unsuitable for a home environment, in which users want to communicate with a speaker across the room. The goal of this study is to propose a novel and robust feature for passive liveness detection. 

\section{Array Fingerprint}
\label{sec:motivation}

In this section, we propose a novel and robust liveness feature \textit{\textbf{array fingerprint}} and elaborate the rationale behind \frameName by answering the following critical questions:

\textbf{RQ1:} How can we model the sound propagation in smart speaker scenarios and answer why existing features (\eg, fieldprint) cannot be effective in such scenarios?

\textbf{RQ2:} How can we extract a useful feature from multi-channel voice samples that is robust regarding a user's location and microphone array's layout?

\textbf{RQ3:} 
What are the benefits of the array fingerprint? Is it effective and robust to the distortions caused by environmental factors?

\subsection{Theoretical Analysis on Sound Propagation for Smart Speakers}
\label{sec:array:rq1}

To answer question \textbf{RQ1}, we give a theoretical analysis of sound propagation in a smart speaker scenario by following the model proposed in Section~\ref{sec:sound:generation} and discuss the limitations of the previous works.

\begin{figure}[t]
\centering
\includegraphics[height=2.6cm]{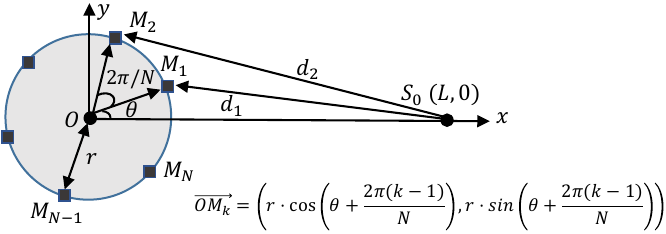}
\caption{Sound propagation in microphone array scenario.}
\vspace{-4mm}
\label{fig:principle::array}
\end{figure}

\noindent \textbf{Sound propagation model for smart speakers.} Figure ~\ref{fig:principle::array} illustrates the scenario when audio signals are transmitted from source to microphone array. The audio source is regarded as a point with coordinate $(L,0)$ and the microphones are evenly distributed on a circle. Given the $k$-th microphone $M_k$, the collected audio data is $y_k(f,t) = h_{air}(d_k,f,t) \cdot s(f,t)$, where $d_k$ is the path distance from the audio source to $M_k$. 
In the theoretical analysis, to simplify the description of the channel gain $h_{air}(d_k,f,t)$, we apply the classic spherical sound 
wave transmission model in air~\cite{hansen2001fundamentals}.\footnote{In real-world scenarios, sound decay in the air is correlated with many factors such as temperature, medium, and surrounding objects. Using the classical model simplifies the question and the effectiveness of \frameName will be demonstrated by experiments in Section~\ref{sec:array:rq3}.} Thus, $h_{air}(d_k, f, t)$ can be estimated as:
\begin{equation}
    h_{air}(d_k, f,t) = Ce^{-\alpha_c d_k} = Ce^{-\alpha (s(f, t)) d_k},
\end{equation}
where C is the attenuation coefficient, and $\alpha_c$ is the absorption coefficient which varies with the signal frequency $f$.  Therefore, we replace $\alpha_c$ with $\alpha(s(f,t))$.
Then, from Section~\ref{sec:sound:generation}, the collected audio in $M_k$ can be represented as:

\begin{equation}
    y_k(f,t) = h_{air}(d_k, f, t)\cdot s(f,t) = Ce^{-\alpha (s(f, t)) d_k}\cdot s(f,t).
    \label{eq:yi}
\end{equation}

\noindent\textbf{Existing passive liveness detection schemes are vulnerable to environmental changes.}
From equation \ref{eq:yi}, it is observed that changing the relative distance between the microphone and audio source will cause non-linear distortion on the microphone's collected signal. 
Such distortion is related to the original $s(f,t)$ and thus is hard to be eliminated. This is the reason why mono channel-based detection schemes are fragile to the change of propagation path.

For the fieldprint-based solution, from equation \ref{eq:field-print}, the extracted feature can be represented as $log(y_i/y_j) = -\alpha(s(f,t))\cdot lg(e) \cdot (d_i - d_j)$. When the positions of the microphone pair are fixed (\ie, $d_i - d_j)$ can be regarded as a constant), the above feature is a function of originally generated $s(f,t)$ containing liveness factor as described in Section~\ref{sec:passive:liveness}. However, when the microphone's position changes, the $d_i - d_j$ will no longer be a stable value, and leveraging such a feature becomes infeasible.

\subsection{Advantage of Array Fingerprint: Definition and Simulation-based Demonstration}
\label{sec:array:rq2}

In this subsection, we answer \textbf{RQ2} by defining the array fingerprint and mathematically demonstrating its effectiveness.

From the theoretical analysis in Section~\ref{sec:array:rq1}, to achieve robust liveness detection, the extracted channel feature has to minimize the effects of the propagation factors such as $C$ and $d_k$. Inspired by the circular layout of microphones in smart speaker as shown in Figure~\ref{fig:principle::array}, we define the array fingerprint $A_F$ as below:
\begin{equation}
\begin{aligned}
    A_F &= std(log[y_1, y_2, ..., y_N])\\
    &= std(C - \alpha(s(f,t))\cdot lg(e)\cdot [d_1, d_2, ..., d_N])\\
    &= -\alpha (s(f,t))\cdot lg(e)\cdot std([d_1, d_2, ..., d_k])\\
    &= A_F(s(f,t), \sigma_d).
\end{aligned}
\label{eq:array}
\end{equation}

From equation \ref{eq:array}, we know that the array fingerprint is mainly dominated by source audio $s(f,t)$ and standard deviation of propagation distances $\sigma_d = std([d1, d2, ..., d_N])$.
However, to effectively capture the audio's identity, which can be derived from $s(f,t)$, the hypothesis that $\sigma_d$ could be regarded as a constant parameter must be proved.

To demonstrate the above hypothesis, the propagation distance between audio source $S_0$ and each microphone should be precisely determined. 
To achieve this goal, as shown in Figure~\ref{fig:principle::array}, we denote the center of the microphone array of the smart speaker and the audio source (\eg, human or electrical machine) as origin $O$ and $S_0 (L, 0)$ respectively. For the $k$-$th$ microphone $M_k$, its coordinate can be represented as:

\begin{equation}
    \overrightarrow{OM_k} = (r\cdot cos(\theta + \frac{2\pi(k-1)}{N}), ~r\cdot sin(\theta + \frac{2\pi(k-1)}{N})),
\end{equation}
where $r$ is of the radius of the microphone array, $N$ is the number of microphones, and $\theta$ is the angle between $M_1$ and $X$ axis. Thus the distance $d_k$ between $S_0$ and $M_k$ could be represented as:
\begin{equation}
    d_k = |\overrightarrow{M_k S_0}| = r\sqrt{1 + (\frac{l}{r})^2 - 2(\frac{l}{r})cos(\theta + \frac{2\pi (k-1)}{N})}.
\end{equation}

\begin{figure}[t]

\centering

\includegraphics[width=7.7cm]{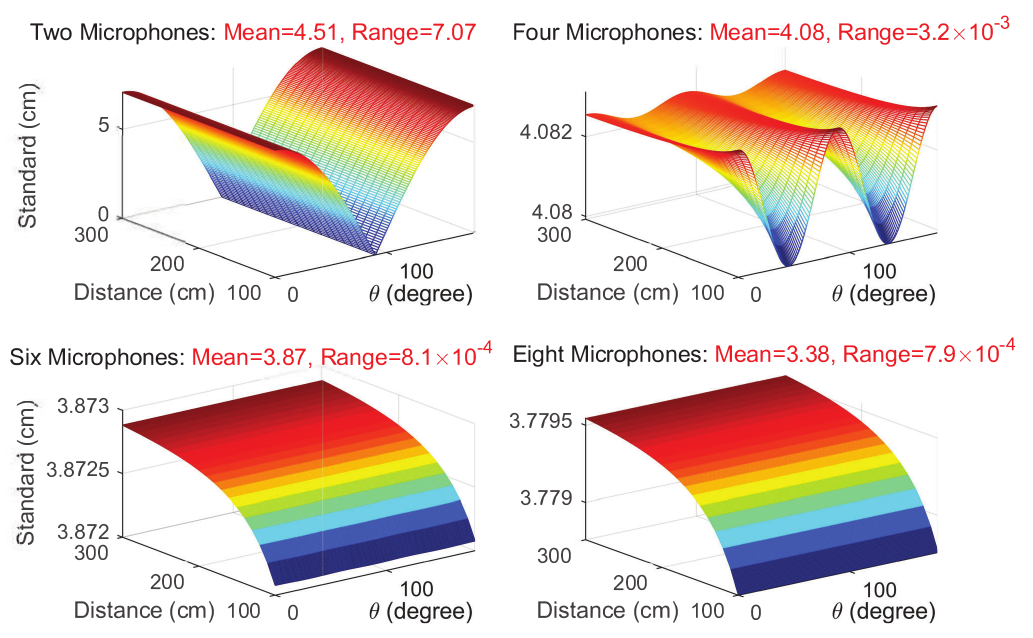}

\caption{$\sigma_d$ values when propagation path changes.}
\vspace{-4mm}

\label{fig:principle:degree}

\end{figure}

To verify the robustness of $\sigma_d$, we performed a simulation based on the multi-microphone speaker with a radius of 5 cm used by Amazon Echo 3rd Gen~\cite{alexa:review}. The distance $L$ varies from 1 m to 3 m, the $\theta$ changes from 0 to 90 degrees. The microphone number $N$ are set as 8, 6, 4, and 2, respectively.

Figure \ref{fig:principle:degree} shows the simulation results under different microphone numbers. When employing more than 4 microphones, the $\sigma_d$ converges to a constant value. For instance, when $N=6$, $\sigma_d$ has an average of 3.38 cm with the range of only $7.9\times10^{-4}$ cm. However, when $N$ is set to 2 (\ie, the scenario in fieldprint-based scheme~\cite{fieldprint2019}), the $\sigma_d$ varies from 0 to 7.07 cm. Since the microphone array of the smart speaker usually has more than four microphones, the $\sigma_d$ 
which is almost unchanged can be regarded as a constant parameter that merely impacts the $A_F$.

From the above theoretical analysis and simulation, it can be derived that the array fingerprint is mainly related to the source audio $s(f,t)$ and thus resilient to the changes of environmental factors, especially for the distance. This is why array fingerprint outperforms other features from mono or two-channel audios~\cite{fieldprint2019, void2020, blue2018hello}.

\begin{figure*}[t]
\centering
\subfigure[Two original authentic audios.]{
\includegraphics[height=2.7cm]{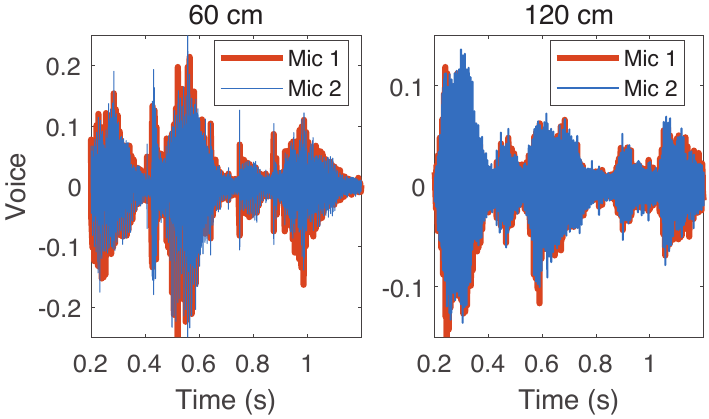}
\label{fig:motivation:original}
}
\subfigure[Dynamic power differences in different microphone pairs.]{
\includegraphics[height=2.7cm]{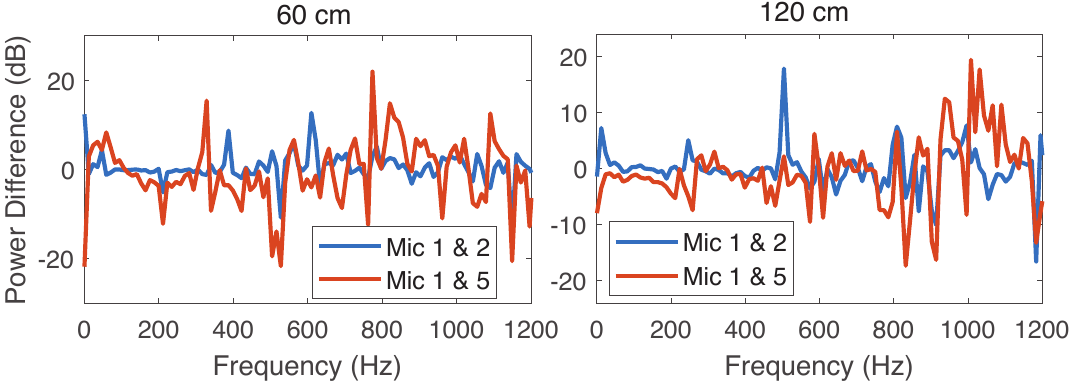}
\label{fig:motivation:fieldprint}
}
\subfigure[Stable array fingerprints.]{
\includegraphics[height=2.7cm]{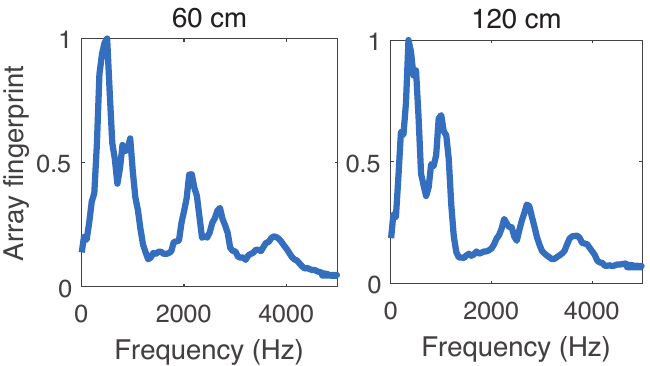}
\label{fig:motivation:arrayid}
}
\vspace{-2mm}
\caption{Illustration of stability of array fingerprint under two locations.}
\vspace{-2mm}
\label{fig:motivation:standard}
\end{figure*}

\begin{figure}
\centering
\includegraphics[width=\linewidth]{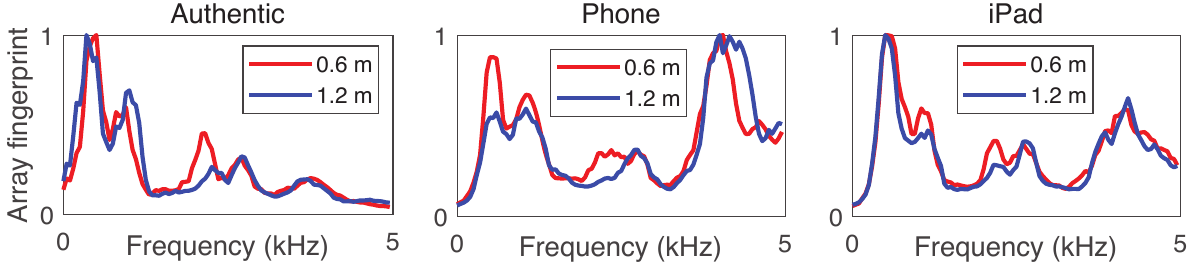}
\vspace{-4mm}
\caption{Differentiating human voice from two spoofing devices via array fingerprints under different propagation paths.}
\vspace{-4mm}
\label{fig:case:study}
\end{figure}

\subsection{Validation of Array Fingerprint}
\label{sec:array:rq3}

Besides theoretical analysis, to answer \textbf{RQ3}, we further validate the effectiveness of the proposed array fingerprint via a series of real-world case studies.

In the experiment, the participant is required to speak the command ``Ok Google" at distances of 0.6 m and 1.2 m, respectively. Figure~\ref{fig:motivation:original} shows the audio signal clips collected by a microphone array with six microphones, and the audio difference between different channels is obvious. When employing the concept of fieldprint, it is observed from Figure~\ref{fig:motivation:fieldprint} that the fieldprints extracted from microphone pair $(M_1, M_2)$ and $(M_1, M_5)$ are quite different.\footnote{The real process of extracting fieldprint is more complicated. Figure~\ref{fig:motivation:fieldprint} shows the basic principle following the descriptions in equation~\ref{eq:field-print}.} Among different distances, the fieldprints are also quite different. However, from Figure~\ref{fig:motivation:arrayid} we can see that the array fingerprints for different distances are very similar.\footnote{This array fingerprint is refined after extracting from equation 4. The detailed calculation steps are described in Section~\ref{sec:feature:sap}.} 

To show the distinctiveness of array-print, we also conducted replay attacks via smartphones and iPad (\ie, device \#8 and \# 3 in Table~\ref{table:device} of Appendix~\ref{appendix:dataset}). The normalized array fingerprints (\ie, $F_{SAP}$ in Section~\ref{sec:feature:sap}) are shown in Figure~\ref{fig:case:study}. It is observed that the array fingerprints for the same audio sources are quite similar, while array fingerprints for different audio sources are quite different. 
Our theoretical analysis and experimental results demonstrate the array fingerprint can serve as a better passive liveness detection feature. This motivates us to design a novel, lightweight and robust system which will be presented in the next section.

\section{The Design of \frameName}
\label{sec:design}

\begin{figure}[t]
\centering
\includegraphics[width=0.85\linewidth]{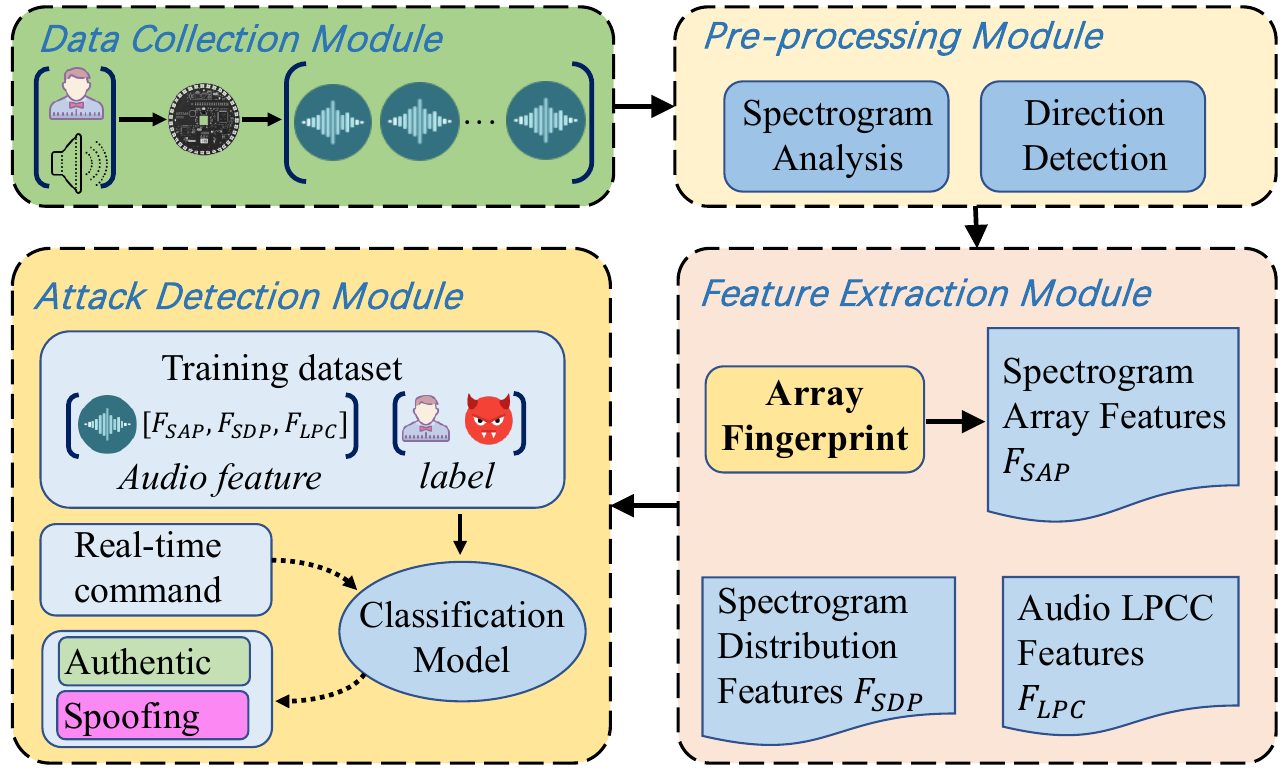}
\caption{System overflow.}
\vspace{-4mm}
\label{fig:overflow}
\end{figure}

As shown in Figure~\ref{fig:overflow}, we propose \frameName, a robust liveness detection system based on the proposed array fingerprint with other auxiliary features. 
\frameName consists of the following modules: \textit{Data Collection Module}, \textit{Pre-processing Module},  \textit{Feature Extraction Module}, and \textit{Attack Detection Module}. We will elaborate on the details of each module in this section.

\begin{figure*}[t]
\center
\subfigure[Original spectrograms of different channels.]{
\includegraphics[height = 2.45cm]{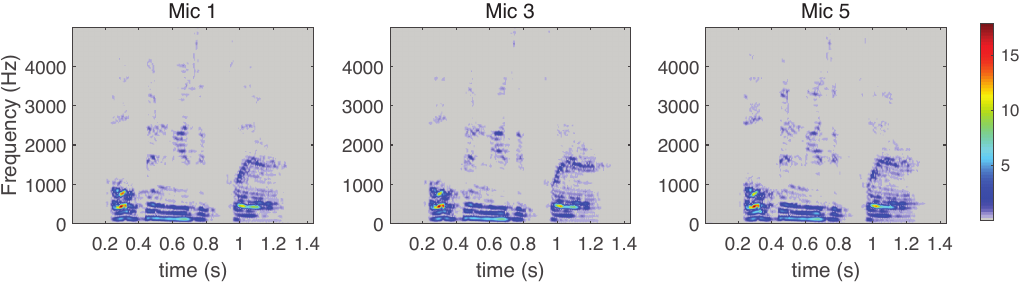}
\label{fig:sap:original}
}
\subfigure[Spectrogram grids of different channels.]{
\includegraphics[height = 2.45cm]{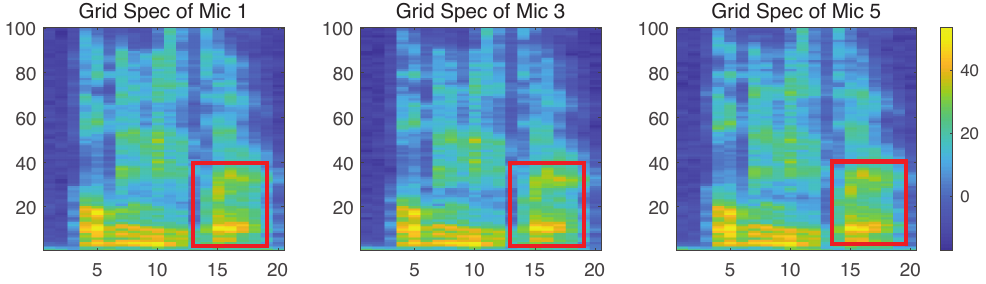}
\label{fig:sap:grid}
}
\subfigure[Array fingerprint Extraction Processing.]{
\includegraphics[height = 2.3cm]{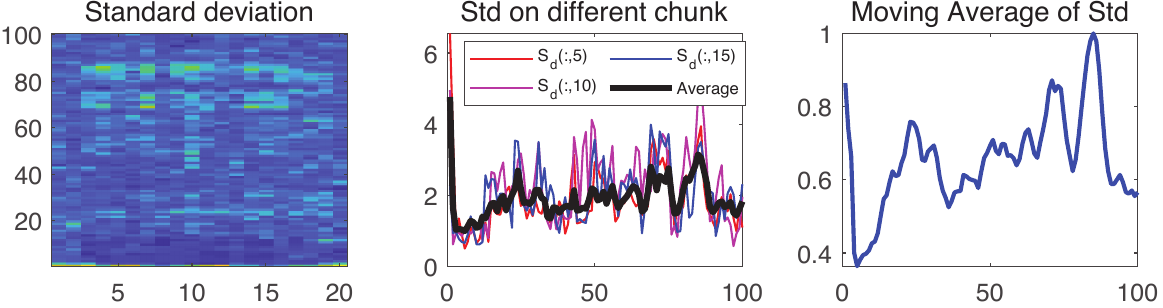}
\label{fig:sap:process}
}
\subfigure[Features among different commands and distances.]{
\includegraphics[height = 2.3cm]{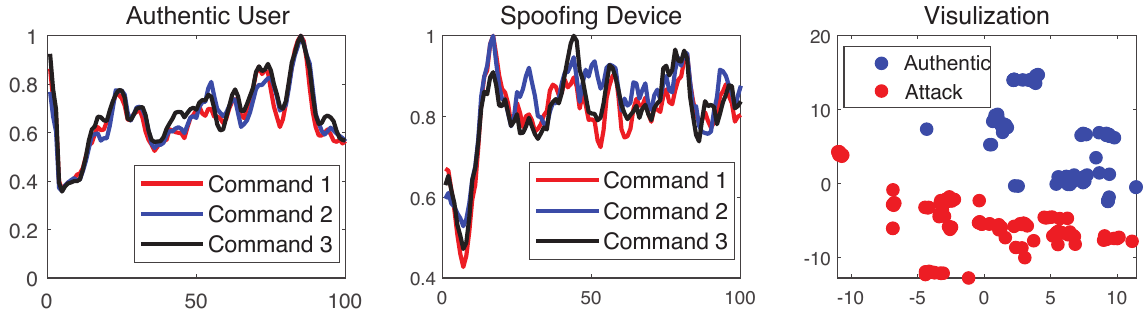}
\label{fig:sap:result}
}
\vspace{-4mm}
\caption{Illustration of spectrogram array fingerprint feature $F_{SAP}$ extraction.} 
\vspace{-4mm}
\label{fig:sap}
\end{figure*}

\subsection{Multi-channel Data Collection}

Currently, most popular smart speakers, such as Amazon Echo and Google, employ a built-in microphone array to collect voice audio. However, due to privacy and commercial concerns, the user of the smart speaker cannot access the original audio data, only the transcribed text.
To solve this problem, we utilize open modular development boards with voice interfaces (\ie, the Matrix Creator~\cite{matrix} and Seeed Respeaker~\cite{respeaker}) to collect the data. Since these development boards have similar sizes to commercial smart speakers, \frameName evaluations on the above devices can be applied to a smart speaker without any notable alterations. Generally speaking, given a smart speaker with $N$ microphones, a sampling rate of $F_s$, and data collection time $T$, the collected voice sample is denoted as $V_{M\times N}$, where $M=F_s\times T$ and we let $V_i$ be the \textit{i-th} channel's audio $V(:,i)$. Then, the collected $V$ is sent to the next module.

\subsection{Data Pre-processing}
\label{sec:data:preprocessing}

As shown in equation~\ref{eq:array}, the identity (\ie, real human or spoofing device) is hidden in the audio's spectrogram. 
Therefore, before feature extraction, we conduct the frequency analysis on each channel's signal and detect the audio's direction.

\noindent \textbf{Frequency analysis on multi-channel audio data.} 
As described in Section~\ref{sec:array:rq2}, the audio spectrogram in the time-frequency domain contains crucial features for further liveness detection. \frameName performs Short-Time Fourier Transform (STFT) to obtain two-dimensional spectrograms of each channel's audio signal. For the $i$-th channel's audio $V_i$, which contains $M$ samples, \frameName applies a Hanning window to divide the signals into small chunks with lengths of 1024 points and overlapping sizes of 728 points. Finally, a 4096-sized Fast Fourier Transform (FFT) is performed for each chunk and a spectrogram $S_i$ is obtained as shown in Figure~\ref{fig:sap:original}.

\noindent \textbf{Direction detection.} 
Given a collected audio $V_{M \times N}$, to determine the microphone which is closest to the audio source, \frameName firstly applies a high pass filter with a cutoff frequency of 100 Hz to $V_{M \times N}$. Then, for the $i$-th microphone $M_i$, \frameName calculates the alignment errors $E_i = mean((V(:, i-1)- V(:, i))^2)$~\cite{direction}. Finally, from the calculated $E$, \frameName chooses the microphone with minimum alignment error as the corresponding microphone.

\subsection{Feature Extraction}
\label{sec:feature}
From
equations~\ref{eq:yi} and~\ref{eq:array}, we observe that both audio spectrograms themselves and the microphone array's difference contain the liveness features of collected audio. In this module, the following three features are selected by \frameName: \textit{Spectrogram Array Fingerprint} $F_{SAP}$, \textit{Spectrogram Distribution Fingerprint} $F_{SDP}$, and \textit{Channel LPCC Features} $F_{LPC}$. 

\subsubsection{Spectrogram Array Feature}
\label{sec:feature:sap}
After obtaining the spectrogram $\textbf{S} = [S_1, S_2, \dots, S_N]$ from $N$ channels' audio data $\textbf{V} = [V_1, V_2, ..., V_N]$, \frameName firstly exploits the array fingerprint which is proposed in Section~\ref{sec:array:rq2} to extract the identity of the audio source. To reduce the computation overhead, for $S_k$ with size $M_s \times N_s$, we only preserve the components in which frequency is less than the cutoff frequency $f_{sap}$. In this study, we empirically set $f_{sap}$ as 5 kHz. The resized spectrograms are denoted as $\textbf{Spec} = [Spec_1, Spec_2, ..., Spec_k]$, where $Spec_k = S_k(:M_{spec}, :)$. In this study, with sampling rate $Fs$ = 48kHz and FFT points $N_{fft} = 4096$, the $M_{spec}$ is $[\frac{f_{sap} \times N_{fft}}{F_s}] = 426$.

\begin{figure*}[t]
\centering
\subfigure[The authentic audio's $Ch$.]{
\includegraphics[height = 1.85cm]{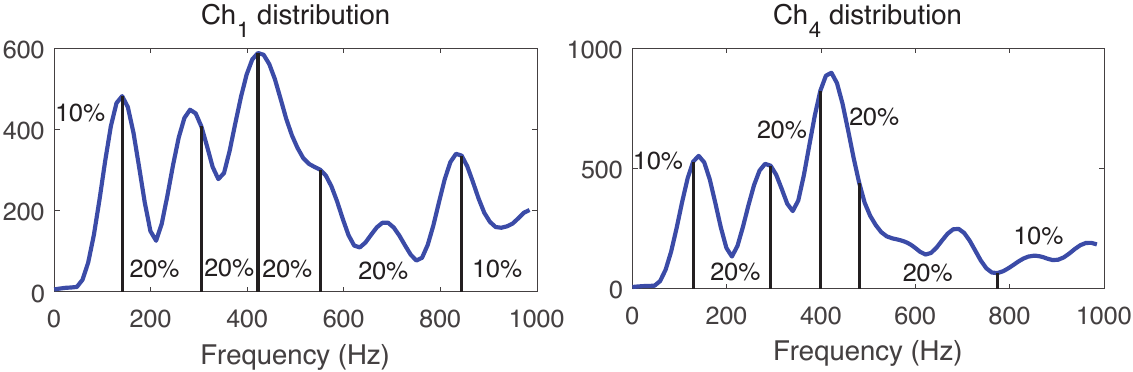}
\label{fig:spd:normal}
}
\subfigure[The spoofing audio's $Ch$.]{
\includegraphics[height = 1.85cm]{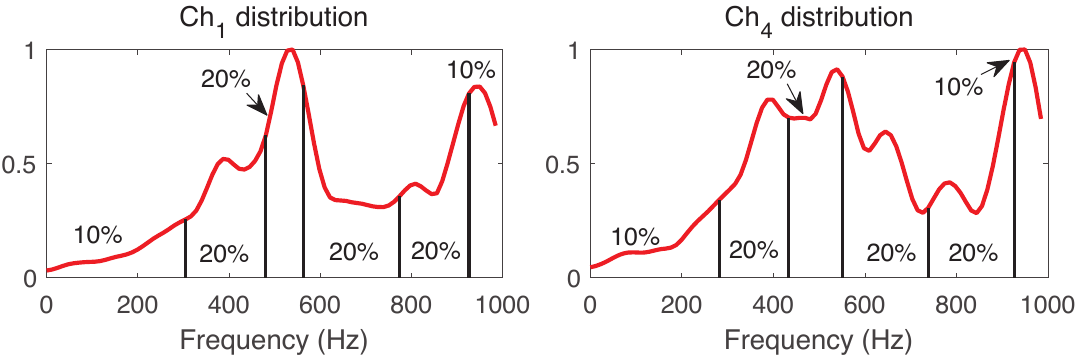}
\label{fig:spd:spoofing}
}
\subfigure[$F_{SDP}$ between authentic and spoofing audios.]{
\includegraphics[height = 1.85cm]{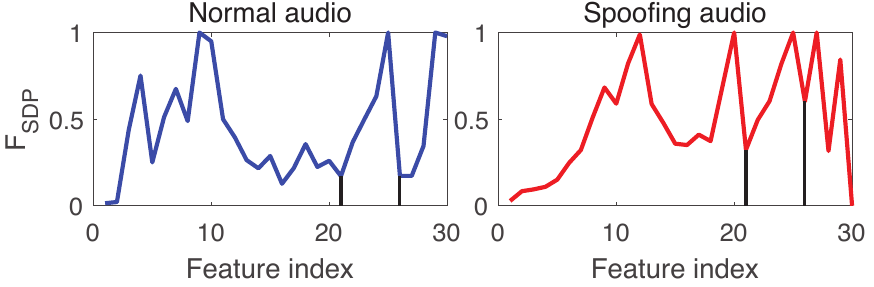}
\label{fig:spd:result}
}
\vspace{-4mm}
\caption{Spectrogram distributions between authentic human and spoofing device.} 
\vspace{-4mm}
\label{fig:spd}
\end{figure*}

Figure~\ref{fig:sap:original} illustrates $Spec$ of three channels of the command "OK Google." It is observed that different channels' spectrograms are slightly different. However, directly using such subtle differences would cause an inaccurate feature. Thus, \frameName transforms $Spec_k$ into a grid matrix $G_k$ with size $M_G \times N_G$ by dividing $Spec_k$ into $M_G \times N_G$ chunks and calculates the sum of magnitudes within each chunk. The element of $G_k$ could be represented as:
\begin{equation}
\begin{split}
    G_k(i,j) = sum( Spec_k(1+(i-1) \cdot S_M: i \cdot S_M, \\
    1+(j-1)\cdot S_N: j\cdot S_N)), \\
\end{split}
\label{equ:grid}
\end{equation}
where $S_M = [\frac{M_{spec}}{M_G}]$ and $S_N = [\frac{N_{spec}}{N_G}]$ are the width and length of each chunk. Note that some elements of $Spec_k$ may be discarded, however, it does not affect the feature generation, since \frameName focuses on the differences between spectrograms according to equation~\ref{eq:array}. In this study, $M_G$ and $N_G$ are set to 100 and 20 respectively, and Figure~\ref{fig:sap:grid} shows the spectrogram grids from the first, third and fifth microphone. The difference among elements in $\textbf{G} = [G_1, G_2, ..., G_N]$ is now very obvious. For instance, the grid values in the red rectangles of Figure~\ref{fig:sap:grid} are quite different.

Then, based on equation~\ref{eq:array}, \frameName calculates the array fingerprint $F_G$ from the spectrogram $\textbf{G}$. $F_G$ has the same size as $G_k$, and the elements of $F_G$ can be represented as:
\begin{equation}
    F_G(i,j) = std([G_1(i,j), G_2(i,j), ..., G_N(i,j)]).
\end{equation}

Figure~\ref{fig:sap:process} illustrates the $F_G$ containing $N_G$ chunks calculated from spectrogram grids as shown in Figure~\ref{fig:sap:grid}. However, we find that in different time chunks, the $F_G(:,i)$ varies. The reason is that different phonemes are pronounced by different articulatory gestures, which can be mapped to a different $h_{user}(f,t)$ function in Section~\ref{sec:sound:generation}. To solve this problem, we leverage the idea that even though different phonemes contain different gestures, there are common components over a long duration of time. Therefore, \frameName averages the $F_G$ across the time axis, and Figure~\ref{fig:sap:process} shows the average result $\overline{F_G}$. \frameName performs a 5-point moving average and normalization on $\overline{F_G}$ to remove noise and generate the spectrogram array fingerprint $F_{SAP}$.

Figure~\ref{fig:sap:result} gives a simple demonstration about the effectiveness of the $F_{SAP}$ feature generation process. We test three voice commands ``OK Google'', ``Turn on Bluetooth'' and ``Record a video'', while the distances between the speaker and microphone array are set as 0.6 m and 1.2 m in the first two commands and the last command, respectively. In Figure~\ref{fig:sap:result}, it is observed that the different commands result in a similar array fingerprint, and the feature difference between authentic audio and spoofing audio is clear. Finally, since \frameName requires a fast response time, the feature should be lightweight. So, the $F_{SAP}$ is re-sampled to the length of $N_{SAP}$ points. In this study, we empirically choose $N_{SAP}$ as 40.

\subsubsection{Spectrogram Distribution Feature}

Besides $F_{SAP}$, as mentioned in equation~\ref{eq:yi}, the spectrogram distribution also provides useful information related to the identity of the audio source. Thus we also extract spectrogram distribution fingerprint $F_{SDP}$ for liveness detection.
Given a spectrogram $S_k$ from the $k$-th channel, \frameName calculates a $N_G$-dimension vector $Ch_k$ in which $Ch_k(i) = \sum_{j = 1}^{M_{spec}}{S_k(j, i)}$, where  $M_{spec}$ and $N_G$ are set as 85 and 20 respectively in this study.\footnote{When calculating $F_{SDP}$, we set the cutoff frequency as 1 kHz since most human voice frequency components are located in the 0\textasciitilde1 kHz range and the corresponding $M_{Spec}$ is 85 under the the parameters in Section~\ref{sec:feature:sap}.} For the audio with $N$ channels, the channel frequency strength  ${Ch} = [Ch_1, Ch_2, ... , Ch_N]$ is obtained.

Figure~\ref{fig:spd:normal} and \ref{fig:spd:spoofing} show channel frequency strength $Ch_1$ and $Ch_4$ of first and fourth channels from authentic and spoofing audios. It is observed that $Ch$ from real human and spoofing device are quite different. Therefore, we utilize the average of channel frequency strengths $\overline{Ch}$ and re-sample its length to $N_{Ch}$ as the first component of $F_{SDP}$. In this study, $\overline{Ch(i)} = mean([Ch_1(i), Ch_2(i), ... , Ch_N(i)])$ and $N_{Ch}$ is set to 20.
We can also find that for the same audio, $Ch$ from different channels have slightly different magnitudes and distributions. To characterize the distribution of $Ch$, for $Ch_k$ from the $k$-th channel, \frameName first calculates the cumulative distribution function $Cum_k$ and then determines the indices $\mu$ which can split $Cum_k$ uniformly. As shown in Figure~\ref{fig:spd:normal} and \ref{fig:spd:spoofing}, the $Ch_k$ are segmented into 6 bands. \frameName sets the $Thr = [0.1, 0.3, 0.5, 0.7, 0.9]$, and the index $\mu(k,i)$ of the i-th $Thr$ for $Ch_k$  satisfies the following condition:

\begin{equation}
    Cum_k(\mu(k, i) \leq Thr_{i} \leq Cum_{k}(\mu(k, i) + 1).
\end{equation}

After obtaining  the $N \times 5$ indices $\mu$, we utilize the mean value $D_{mean}$ and standard deviation $D_{std}$ among different channels as a part of the spectrogram feature. Both $D_{mean}$ and $D_{std}$ are vectors with length of 5, where $D_{mean}(i) = mean(\mu(:,i))$ and $D_{std}(i) = std(\mu(:,i))$.
Finally, \frameName obtains the spectrogram distribution fingerprint $F_{SDP} = [\overline{Ch}, D_{mean}, D_{std}]$. Figure~\ref{fig:spd:result} illustrates the $F_{SDP}$ from authentic and spoofing audios and demonstrates the robustness of $F_{SDP}$.

\subsubsection{Channel LPCC Features}
\label{sec:feature:lpc}

The final feature of \frameName is the Linear Prediction Cepstrum Coefficients (LPCC). 
Since each channel has unique physical properties, retaining the LPCC which characterizes a given audio signal could further improve the detection performance. For audio signal $y_k(t)$ collected by microphone $M_k$, \frameName calculates the LPCC with the order $p=15$. Due to page limit, the details of LPCC extraction
is introduced in Figure~\ref{fig:lpcc} and Appendix~\ref{appendix:lpcc} respectively.
To reduce the time overhead spent on LPCC extraction, we only preserve the LPCCs from aduios in these two channels $(M_i, M_{mod(i + N/2, N)})$, where $M_{i}$ is the closet microphone derived from Section~\ref{sec:data:preprocessing}.
Finally, we generate the final feature vector $X = [F_{SAP}, F_{SDP}, F_{LPC}]$.

\subsection{Classification Model}
\label{sec:model}
After generating the feature vector from the audio input, we choose a lightweight feed-forward back-propagation neural network to perform liveness detection. The neural network only contains three hidden layers with rectified-linear activation (layer sizes: 64, 32, 16). 
We adopt a lightweight neural network because it can achieve a quick response to the decision, which is essential for the devices in the smart home environment.
We also discuss other possible classification models in Appendix~\ref{appendix:comparison}.

\section{Evaluations}
\label{sec:evaluation}

\subsection{Experiment Setup} 
\label{sec:eval:experiment}

\textbf{Hardware setup.}
Since it is hard for users to obtain audio files from popular smart speakers such as Google Home and Amazon Echo, in this study, to collect multi-channel audios, as shown in Figure~\ref{fig:devices} of Appendix~\ref{appendix:dataset}, we employ two open modular development boards (\ie, Matrix Creator, and Seeed ReSpeaker Core v2) with the sampling rate of 48 kHz to serve as smart speakers. The number of microphones in the Matrix and ReSpeaker are 8 and 6, respectively, and their radiuses are 5.4 cm and 4.7 cm respectively. For the spoofing device, we employ 14 different electrical devices with various sizes and audio qualities whose detailed parameters are shown in Table~\ref{table:device} of Appendix~\ref{appendix:dataset}.

\noindent \textbf{Data collection procedure.}
In this study, 20 participants are recruited to provide the multi-channel audio data. 
The data collection procedure consists of two phases: \textit{(\romannumeral1) Authentic audio collection:} 
in this phase, the participant speaks 20 different voice commands as listed in Appendix~\ref{appendix:dataset} and the experimental session can be repeated multiple times by this participant. We pre-define four distances (\ie, 0.6 m, 1.2 m, 1.8 m, 2.4 m) between the microphone array and the participant can choose any of them in each session. For the speaking behavior, we ask the participant to speak command as she/he likes and did not specify any fixed speed/tone.  
\textit{(\romannumeral2) Spoofing audio collection:} in this phase, similar to the manners adopted by the previous works~\cite{fieldprint2019,void2020,viblive}, after collecting the authentic voice samples, we utilize the spoofing devices as listed in Table~\ref{table:device} to automatically replay the samples without the participant's involvement. When replaying a voice command, the electrical device is placed at the same location as the corresponding participant.

\noindent \textbf{Dataset description.}
After finishing experiments, we utilize pyAudioAnalysis tool to split the collected audio into multiple voice command samples.\footnote{PyAudioAnalysis website: https://pypi.org/project/pyAudioAnalysis/.} 
After removing incorrectly recognized samples, we get a dataset containing 32,780 audio samples. 
We refer to this dataset as MALD dataset and utilize it to assess \frameName.\footnote{MALD is the abbreviation of ``microphone array-based liveness detection''.}
The details of MALD dataset are shown in Table~\ref{table:overall} of Appendix~\ref{appendix:dataset}. For instance, user \#7 provides 600 authentic samples at three different positions (\ie, the distance of 0.6 m, 1.2 m and 1.8 m) and we utilize these collected samples with three spoofing devices (\ie, SoundLink, iPad, iPhone) to generate 1,800 spoofing samples.

\noindent {\textbf{Training procedure.}} 
{
As mentioned in Section~\ref{sec:model}, \frameName needs to be trained with audio samples before detecting spoofing attacks. 
When evaluating the overall performance of \frameName on the collected MALD dataset in Section~\ref{sec:eval:performance}, we perform the two-fold cross-validation. In each fold (\ie, training procedure), half samples are chosen to generate a classifier and the validation dataset proportion is set as 30\%. 
When evaluating the impact of other factors as shown in Section~\ref{sec:eval:conditions} and Section~\ref{sec:robustness}, the training procedure depends on the specific experiment, and we show the training dataset before presenting the evaluation results.}

\noindent \textbf{Evaluation metrics.} Similar to previous works~\cite{fieldprint2019,Meng:2018:WES:3209582.3209591, void2020}, in this study, we choose accuracy, false acceptance rate (FAR), false rejection rate (FRR), true rejection rate (TRR), and equal error rate (ERR) as metrics to evaluate \frameName. The accuracy means the percentage of the correctly recognized samples among all samples.
FAR represents the rate at which a spoofing sample is wrongly accepted by \frameName, and FRR characterizes the rate at which an authentic sample is falsely rejected. 
EER provides a balanced view of FAR and FRR and it is the rate at which the FAR is equal to FRR. 

\noindent \textbf{Ethics consideration.}
The experiments are under the approval of the institutional review board (IRB) of our institutions. During the experiments, we explicitly inform the participants about the experimental purpose. Since only the voice data are collected and stored in an encrypted dataset, there is no health or privacy risk for the participant.

\subsection{Performance of \frameName}
\label{sec:eval:performance}

\noindent \textbf{Overall accuracy.}
When evaluating \frameName on our own {MALD dataset}, we choose two-fold cross-validation, which means the training and testing datasets are divided equally. 
\frameName achieves the detection accuracy of 99.84\% and the EER of 0.17\%. More specifically, for all 32,780 samples, the overall FAR and FRR are only 0.05\% (\ie, 13 out of 22,539 spoofing samples are wrongly accepted) and 0.39\% (\ie, 40 out of 10,241 authentic samples are wrongly rejected) respectively. The results show that \frameName is highly effective in thwarting spoofing attacks.

\begin{figure}
\centering
\includegraphics[width=0.95\linewidth]{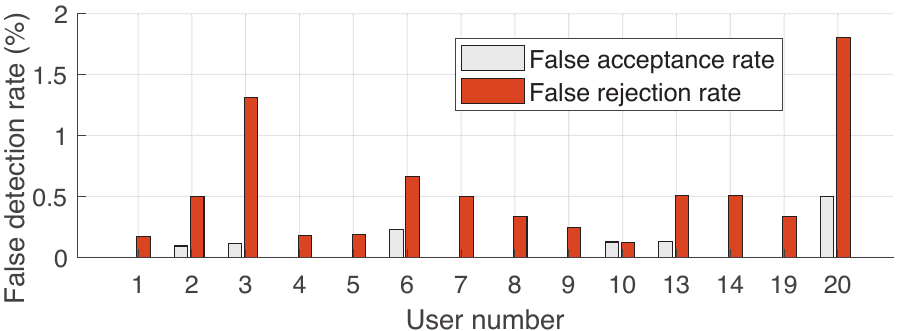}
\caption{Per-user breakdown analysis.}
\vspace{-4mm}
\label{fig:per-user-breakdown}
\end{figure}

\noindent \textbf{Per-user breakdown analysis.}
To evaluate the performance of \frameName on different users, we show the FAR and FRR of each user in Figure~\ref{fig:per-user-breakdown}. Note that, for six users (\ie, users \#11, \#12, \#15, \#16, \#17, \#18) which are not shown in this figure, there is no detection error. When considering FAR, it is observed that the false acceptance cases only exist in 6 users. Even in the worst cases (\ie, user \#20), the false acceptance rate is still less than 0.51\%. When considering FRR, the false rejection cases are distributed among 14 users. It's observed that only the FRRs of users \#3 and \#20 are above 1\%. Although the performance of \frameName on different users is different, even for the worst-case (\ie, user \#20), the detection accuracy is still at 99.0\%, which demonstrates the effectiveness of \frameName.

\noindent \textbf{Time overhead.}
For a desktop with Intel i7-7700T CPU and 16 GB RAM, the average time overhead on 6-channel and 8-channel audios are 0.12 second and 0.38 seconds, respectively. 
Note that it is easy for the existing smart home systems(\eg, Amazon Alexa)  to incorporate \frameName  to their current industrial level solutions in the near future. In that case, both the speech recognition and liveness detection can be done in the cloud \cite{alexa:review}. Therefore, by leveraging the hardware configuration of the smart speaker's cloud (\eg, Amazon Cloud~\cite{amazon:workflow}), which is much better than our existing one (CPU processor), we believe that the time overhead can be further reduced and will not incur notable delays.

\noindent \textbf{Comparison with previous works.}
We further compare the performance of \frameName with existing works to demonstrate the superiority of the proposed array fingerprints. 
{
To eliminate the potential bias in our collected MALD dataset, we also exploit a third-party dataset named ReMasc Core which contains 12,023 voice samples. 
from 40 different users.}
\footnote{{We only consider the 12,023 audio samples collected by circular microphone arrays in the ReMasc Core dataset.}} We re-implement mono audio-based scheme \void~\cite{void2020} and two-channel audio-based scheme \cafield~\cite{fieldprint2019}. For a fair comparison, we replicate their parameters and classification models as shown in Appendix~\ref{appendix:comparison}.

{As shown in Table \ref{tab:comparison}, since MALD dataset is collected in the indoor smart home environment and ReMasc is collected in both indoor, outdoor, and vehicle environments, the detection accuracy varies among these two datasets.
\frameName is superior to previous works in both datasets. 
Especially for the ReMasc Core dataset in which only half of the audio samples are collected in the indoor environment, \frameName is the only scheme that achieves an accuracy above 98.25\%.} The two-channel-based scheme \cafield, gets relatively low performance on both the MALD dataset and ReMasc dataset. It is quite natural since \cafield claimed it needs the user to hold the device with fixed gestures and short distances. {In summary, these results demonstrate that compared with mono audio-based or two-channel-based scheme, exploiting microphone array-based feature achieves superior performance in the liveness detection task. }

\begin{table}[t]
\centering
\caption{{The detection accuracy on both datasets.} }
\label{tab:comparison}
\begin{tabular}{c|c|c}
\hline
\multirow{2}{*}{Liveness feature} & \multicolumn{2}{c}{{Dataset}}  \\ \cline{2-3} 
                                  & {MALD dataset} & {ReMasc dataset} \\ \hline
Microphone array         & 99.84\%      & 97.78\%        \\ \hline

Mono feature                      & 98.81\%      & 84.37\%        \\ \hline
Two-channel                       & 77.99\%      & 82.44\%        \\ \hline
\end{tabular}
\vspace{-4mm}
\end{table}

\subsection{Impact of Various Factors on \frameName}
\label{sec:eval:conditions}

In this subsection, we evaluate the impact of various factors (\eg, distance, direction, user movement, spoofing device, microphone array type) on \frameName.

\begin{table}[t]
\caption{Performance when changing the distance.}
\label{tab:different:location}
\centering
\begin{tabular}{c|c|c|c}
\hline
Training position (m) & 1.2        & 1.8        & 2.4        \\ \hline
Accuracy (\%)   & 99.41  & 99.53  & 99.66 \\ \hline
EER (\%)   & 1.11 & 0.93 & 0.69 \\ \hline
\end{tabular}
\end{table}

\noindent \textbf{Impact of changing distance.}
To evaluate the performance of \frameName on a totally new distance, we recruit four participants to attend experiments at three different locations (i.e., 1.2 m, 1.8 m, 2.4 m).
We totally collect 2,410 authentic and 2,379 spoofing audio samples. 
For a given distance, the classifier is trained with audios at this distance and tested on audios at other distances. 
As shown in Table~\ref{tab:different:location}, compared with the performance in Section~\ref{sec:eval:performance},  \frameName's performance undergoes degradation when the audio source (\ie, the human or the spoofing device) changes its location. 
However, in all cases, \frameName achieves an accuracy above 99.4\%, which demonstrates \frameName is robust to the training distance. This result is also conform to the theoretical analysis in Section~\ref{sec:array:rq2}

\begin{table}[t]
\caption{Performance under different directions.}
\centering
\label{table:direction}
\begin{tabular}{c|c|c|c|c}
\hline
Direction               & Front & Left    & Right   & Back \\ \hline
\# authentic samples & 1020  & 1004    & 1195    & 1000     \\ \hline
\# spoofing samples & 980   & 947     & 971     & 932     \\ \hline
Accuracy (\%) & 100 & 99.69 & 99.31 & 99.74 \\ \hline
EER (\%)                & 0   & 0.59  & 1.08  & 0.43  \\ \hline
\end{tabular}
\vspace{-4mm}
\end{table}

\noindent \textbf{Impact of changing direction.}
In Section~\ref{sec:eval:experiment}, when collecting audio samples, most participants face the smart speaker while generating voice commands. To explore the impact of the angles between the user's face direction and the microphone array, we recruit 10 participants to additionally collect authentic voice samples in four different directions (\ie, front, left, right, back) and then the spoofing device \#8 in Table~\ref{table:device} is utilized to generate spoofing audios. As shown in  Table~\ref{table:direction}, we totally collect 4,219 authentic samples and 3,830 spoofing samples. Then, we use the classification model trained in Section~\ref{sec:eval:performance} to conduct liveness detection. It is observed from Table~\ref{table:direction} that in all scenarios, \frameName achieves an accuracy above 99.3\%, which means \frameName is robust to the change of direction.

\noindent \textbf{Impact of user movement.}
As similar to the above paragraphs, we recruit 10 participants to speak while walking. Then, the participant walks while holding a spoofing device (i.e., Amazon Echo) and plays spoofing audios. We collect 1,999 authentic and 1,799 spoofing samples, and the classifier is the same as that in Section~\ref{sec:eval:performance}. The detection accuracy is 98.2\% which demonstrates that \frameName and the array fingerprint are robust even with the user's movement.

\noindent \textbf{Impact of changing environment.}
To evaluate the impact of different environments on \frameName, we recruit 10 participants to speak voice commands and use device \#8 to launch voice spoofing at a room different from that in Section~\ref{sec:eval:performance}. We collect 1,988 authentic samples and 1,882 spoofing samples respectively. When utilizing the classifier in Section~\ref{sec:eval:performance}, the detection accuracy is 99.30\%, which shows \frameName can effectively thwart voice spoofing under various environments.

\noindent \textbf{Impact of microphone numbers in the smart speaker.}
Studying the relationship between \frameName's performance and the number of microphones could help the smart speaker vendors to determine microphone configurations. 
Note that the data in {MALD dataset} can be divided into six-channel (collected by ReSpeaker) and eight-channel (collected by Matrix) audios. Then, we generate four-channel audio data from the data collected by Matrix device by extracting data from microphones $(M_1, M_3, M_5, M_7)$. 

For three audio groups with 4, 6, and 8 channels respectively,  after conducting two-fold cross-validation on each group, the detection accuracies of \frameName are 99.78\%, 99.82\%, and 99.90\% respectively. 
That means changing the number of channels doesn't cause a significant effect on \frameName's performance. From the theoretical analysis in Section~\ref{sec:array:rq2} and Figure~\ref{fig:principle:degree}, the standard deviation of paths from source to each microphone could be regarded as a constant in a smart speaker scenario. Therefore, as long as the microphone array has a circular layout, \frameName could provide robust protection on thwarting voice spoofing.

\noindent \textbf{Impact of Spoofing Devices.}
It is well known that different devices have different frequency-amplitude response properties, and thus may have different attacker power. To evaluate \frameName's performance on thwarting different spoofing devices, we conduct an experiment based on the {MALD dataset} containing 14 spoofing devices as listed in Table~\ref{table:device} of Appendix~\ref{appendix:dataset}. As discussed in Section~\ref{sec:discuss:enroll}, to reduce the user's enrollment burden, we set the training proportion as 10\%. 

Table~\ref{tab:device:performance} illustrates the FAR of \frameName on each device in this case. 
It is observed that among 14 devices, the overall FAR is 0.58\% (\ie, 117 out of 20,290 spoofing samples are wrongly accepted). Besides, \frameName achieves overall 
100\% detection accuracy on 5 devices (\ie, devices \#2, \#3, \#5, \#6, \#7).  Even in the worst case (\ie, device \#12 Megaboom), the true rejection rate is still at 95.86\%. Furthermore, as shown in Section~\ref{sec:eval:performance}, when increasing the training proportion to 50\%, the false accept rate (FAR) of \frameName is only 0.05\%. In summary, \frameName is robust to various spoofing devices. 

\begin{table}[t]
\caption{The FAR of each spoofing device.}
\centering
\label{tab:device:performance}
\begin{tabular}{c|c|c|c|c|c}
\hline
Device \# & 1    & 4    & 8    & 9    & 10   \\ \hline
FAR (\%)  & 0.09 & 1.04 & 0.05 & 0.55 & 0.96 \\ \hline
Device \# & 11   & 12   & 13   & 14   & Others                    \\ \hline
FAR (\%)  & 3.15 & 4.14 & 0.79 & 0.76 & 0                         \\ \hline
\end{tabular}
\end{table}

\subsection{Robustness of \frameName}
\label{sec:robustness}

\subsubsection{{Handling the Incomplete Enrollment Prodecure}}
\label{sec:discuss:unseen}

{Similar to previous works \cite{void2020,fieldprint2019,viblive}, in Section~\ref{sec:eval:performance}, \frameName requires the user to participate in the enrollment procedures (\ie, providing both authentic and spoofing voice samples). Considering that participating in enrollment is not always feasible, we explore the robustness of \frameName in handling the case that users who did not participate in the complete enrollment procedures. }

\begin{figure}[t]
\centering
\includegraphics[width=0.85\linewidth]{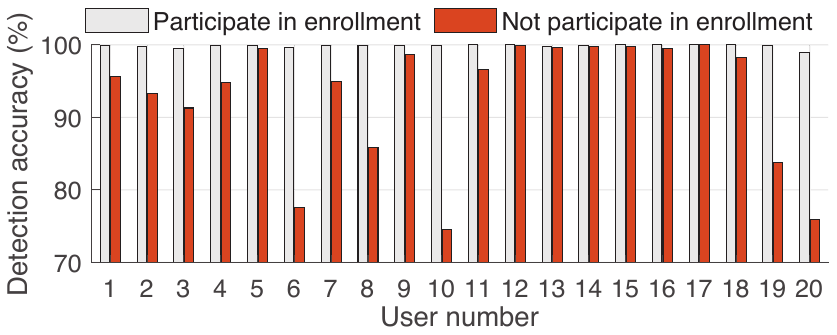}
\caption{{Detection accuracy when the user did not participates in enrollment.}}
\vspace{-4mm}
\label{fig:unseen}
\end{figure}

\noindent
\textbf{Case 1: handling users who did not participate in any enrollment procedure.} In this case, we add an experiment to evaluate the performance of \frameName on participants that did not participate in the enrollment (\ie, \textit{unseen} users).
In the experiment, for each user in the MALD dataset, we train the classifier using other 19 users' legitimate and spoofing voice samples and regard the user's samples as the testing dataset. 
The detection accuracy of each user is shown in Figure~\ref{fig:unseen}. We also show the results described in Section~\ref{sec:eval:performance} when users participate in the enrollment as a comparison.

{
From Figure~\ref{fig:unseen}, it is observed that the overall detection accuracy decreases from 99.84\% to 92.97\%. In the worst case (\ie, user \#12), the detection accuracy decreases from 99.87\% to 74.53\%. The results demonstrate that ability of \frameName on addressing unseen users varies with different users. However, for 11 users, \frameName can still achieve detection accuracies higher than 95\%. The overall results demonstrate that \frameName is still effective when addressing unseen users.
}

{
The performance degradation when addressing unseen users remains an open problem in the area of liveness detection~\cite{void2020,blue2018hello,viblive,Meng:2018:WES:3209582.3209591}. To partially mitigate this issue, a practical solution is requiring the unseen users to provide only authentic voice samples to enhance the classifier (\ie, case 2 discussed below).
}

\noindent
\textbf{{Case 2: handling a user with only authentic samples (without spoofing samples).}}
{
In this case, we consider another situation that the user partially participates in the enrollment and provides only authentic voice samples. 
We add an experiment by leveraging the MALD dataset. Note that, we assume the attacker only utilizes existing devices in the smart home to conduct spoofing. Thus a total of 18 users are selected (\ie, users \#19 and \#20 are excluded because their spoofing devices are never used by others in MALD dataset), whose spoofing devices are listed in Table~\ref{table:device} of Appendix~\ref{appendix:dataset}.
During the experiment, for each selected user, \frameName is trained with this user's authentic voice samples, and generic spoofing samples provided by other 17 users. Then, in the evaluation phase, we test the ability of \frameName to detect attack samples of this user and calculate the corresponding detection accuracy (\ie, TRR).
}

\begin{figure}
\centering
\includegraphics[width=0.95\linewidth]{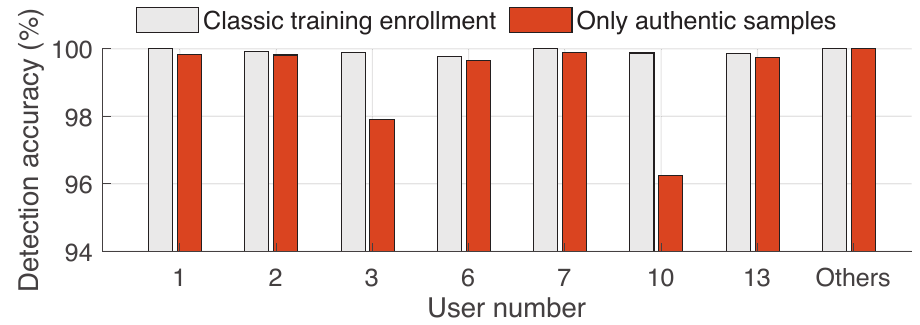}
\caption{{Detection performance under partial enrollment.}}
\vspace{-4mm}
\label{fig:user-only_authentic}
\end{figure}

Figure~\ref{fig:user-only_authentic} illustrates the detection accuracy under two different {enrollment configurations}. For all 18 users, the overall accuracy (\ie, TRR) decreases from 99.96\% in the classical {enrollment scenario} described in Section~\ref{sec:eval:performance} to 99.68\% in this {partial enrollment scenario}. For 11 users (\ie, user \#4, \#5, \#8, \#9, \#11, \#12, \#14, \#15, \#16, \#17, \#18), the accuracy remains 100\% in both scenarios. For the other 7 users, the accuracy decreases slightly due to a lack of knowledge of the user's attack samples in the classifier, but all of them achieve the accuracy of above 96\%, which demonstrates the effectiveness of \frameName in the {partial enrollment scenario.
}

\subsubsection{Liveness Detection on Noisy Environments}
\label{sec:eval:noise}

\begin{figure}[t]
	\centering
	\subfigure[Noise evaluation setting.]{
		\label{fig:noise:setting}
		\includegraphics[width=0.9\linewidth]{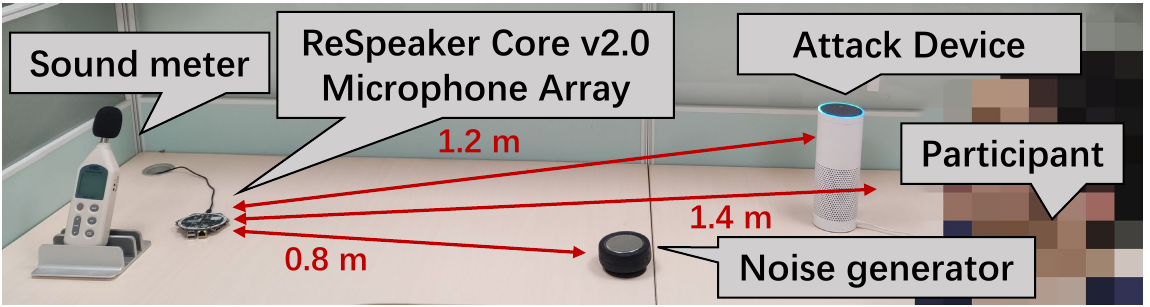}}
	\subfigure[Accuracy and EER.]{
		\label{fig:noise:results}
		\includegraphics[width=0.85\linewidth]{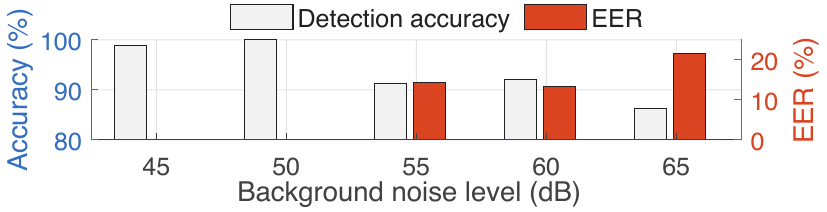}}
	\caption{Performance under noisy environments.}
\vspace{-4mm}
\end{figure}

We add an experiment to evaluate the impact of background noise. As shown in Figure~\ref{fig:noise:setting}, to ensure the noise level is consistent when the user is speaking a voice command, we place a noise generator to play white noise during the data collection. 
We utilize an advanced sound level meter (\ie, Smart Sensor AR824) with an A-weighted decibel to measure the background noise level. 
The strengths of noise level at the microphone array are set as 45 dB, 50 dB, 55 dB, 60 dB, and 65 dB respectively, and a total of 4,528 audio samples are collected from 10 participants and the spoofing device \#13 (\ie, Amazon Echo plus).

We utilize the classifier in Section~\ref{sec:eval:performance} where the noise level is 30 dB to conduct liveness detection. As shown in Figure~\ref{fig:noise:results}, when increasing the noise level from 45 dB to 65 dB, the accuracy decreases from 98.8 \% to 86.3 \%. It is observed that \frameName can still work well when the background noise is less than 50 dB, which also explains why \frameName can handle the audio samples of the ReMasc Core dataset collected in an outdoor environment. However, when there exists strong noise, since the feature of \frameName is only based on the collected audios, the performance of \frameName degrades sharply. We discuss this limitation in Section~\ref{sec:limitations} and leave it for future work.

\subsubsection{Defending against Advanced Spoofing Attacks}
\label{sec:eval:advanced}

\begin{figure}[t]
	\centering
	\subfigure[Spectrums of authentic, replay, and modulated audios.]{
		\label{fig:modulated:spectrum}
		\includegraphics[width=\linewidth]{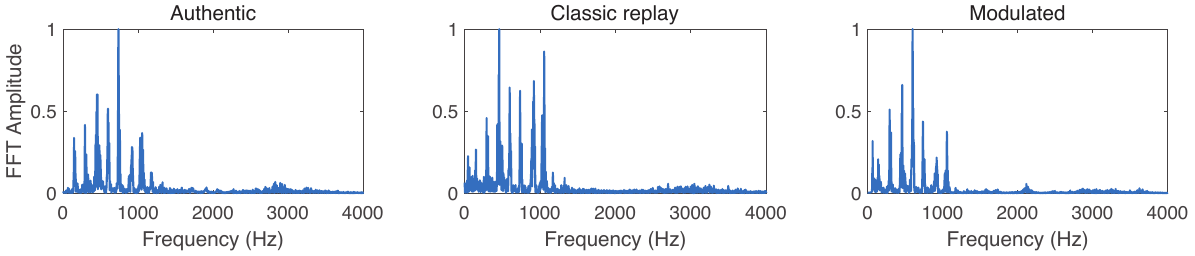}}
	\subfigure[Array fingerprints of authentic, replay, and modulated audios.]{
		\label{fig:modulated:array}
		\includegraphics[width=\linewidth]{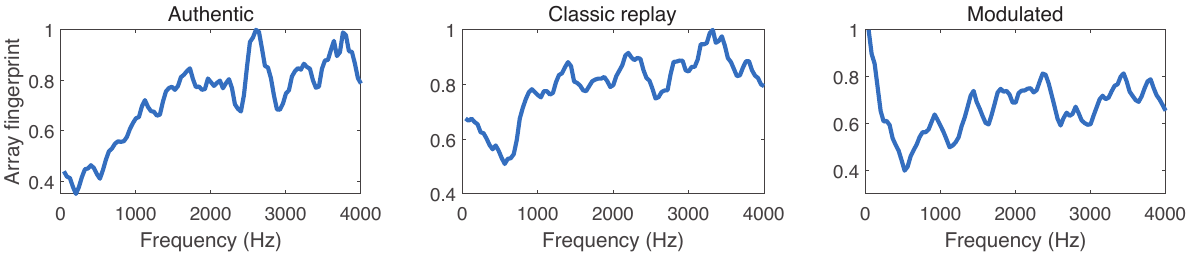}}
	\caption{Spectrums and array fingerprints of audio signals.
}
\vspace{-4mm}
	\label{fig:modulated}
\end{figure}

\noindent {\textbf{Thwarting modulated attacks.}
In this subsection, we first study the performance of \frameName under the emerging modulated attack~\cite{wangccs2020}. By modulating the spectrum of replayed audio, the modulated attack~\cite{wangccs2020} identifies an important threat to existing liveness detection schemes.
To achieve this goal, in the attack model, the adversary first needs to use a microphone of the target device to collect the target user's authentic voice samples.}
\footnote{{The attack assumption of the modulated attack~\cite{wangccs2020} only considers the voice interface with only one microphone.}} 
{Then, the adversary physically approaches the spoofing device to measure its frequency amplitude curve and the corresponding inverse filter using the target microphone. Finally, by applying the inverse filter on the authentic audio and playback it via the spoofing device, for the target microphone, the spectrum of the collected modulated audio is similar to the collected authentic audio as shown in Figure~\ref{fig:modulated:spectrum}. 
However, since the array fingerprint characterizes the difference among the multiple microphones, it is feasible for \frameName to thwart modulated attacks.}

We conduct a case study to demonstrate the robustness of the array fingerprint. We select an Amazon Echo and a ReSpeaker microphone array as the spoofing and target device, respectively, and follow the steps in~\cite{wangccs2020} to re-implement modulated attack. We recruit a volunteer to provide an authentic voice command and then collect its corresponding classic replay and modulated audios generated by the Echo device.

Figure~\ref{fig:modulated} shows spectrums and array fingerprints of authentic audio and its corresponding replay and modulated samples. It is observed from Figure~\ref{fig:modulated:spectrum} that, for a given channel, the spectrum of modulated audio (\ie, FFT Amplitude of the first channel audio $V_1$) is similar to that in the authentic audio, which means it can bypass many existing liveness detection schemes. However, since the human vocal organs and spoofing devices cannot be regarded as a point sound source, the sounds received in multiple microphones show the obvious differences.\footnote{In the theoretical analysis of Section~\ref{sec:array:rq1}, to simplify the analysis of the classic replay attack, we regard the human and loudspeaker as points.} 
And the difference between multiple channel audios (\ie, six channels in this experiment) characterized by array fingerprints still retains the audio's identity. As shown in Figure~\ref{fig:modulated:array}, the array fingerprint of the modulated sample is still similar to that of classic replay audio, which shows it is feasible for \frameName to thwart the modulated attack. 

Then, we evaluate the effectiveness of \frameName on thwarting the modulated attack. 
In the experiment, we choose three different spoofing loudspeakers \#3, \#13, and \#14 (\ie, Echo Plus, iPad 9, and Mi 9). 
We recruit 10 participants to provide authentic samples and follow the steps described in~\cite{wangccs2020} to generate 1,990, 1,791, and 1,994 modulated attack samples for Echo, iPad, and Mi respectively. Due to the page limit, the details of modulated attacks are shown in Appendix~\ref{appendix:inverse}. 

When employing the classifier in Section~\ref{sec:eval:performance}, the accuracy of \frameName on detecting the modulated samples among Echo, iPad, and Mi are 100\%, 92.74\%, and 97.29\% respectively. 
In summary, \frameName can successfully defend against the modulated attack, but the performance varies with different spoofing devices. Considering combining \frameName with the dual-domain detection proposed in~\cite{wangccs2020} can further improve the security of smart speakers.

\noindent \textbf{Other adversarial example attacks.}
To validate \frameName's robustness under adversarial attacks, we re-implement hidden voice attacks~\cite{2016Nicholas} and VMask~\cite{vmask} which breach speech recognition and speaker verification schemes, respectively. For each type of attack, we conduct voice spoofing 100 times, and the experimental results show that \frameName detects 100\% of attack audios for both attacks.
The reason why \frameName could detect these attacks is that these attacks only aim to add subtle noises into source audio to manipulate the features (\eg, MFCC) interested by speech/speaker recognition schemes but the array fingerprint cannot be fully converted to that of the target victim.

\section{Discussions}
\label{sec:discussions}

\subsection{User Enrollment Time in Training}
\label{sec:discuss:enroll}

\noindent \textbf{Impact of training dataset size.}
To reduce the user's registration burden, we explore the impact of training data size on the performance of \frameName. For our collected {MALD dataset}, we set the training dataset proportion as 10\%, 20\%, 30\%, and 50\% respectively. The results are shown in Table~\ref{tab:enrollment}. It is observed that the detection performance increases from 99.14\% to 99.84\% when involving more training samples. Note that, even if we only choose 10\% samples for training, \frameName still achieves the accuracy of 99.14\% and EER of 0.96\%, which is superior to previous works~\cite{viblive}.

\noindent \textbf{Time overhead of user's enrollment.}
As mentioned in Section~\ref{sec:eval:experiment}, the participant does not need to provide spoofing audio samples. Besides, as shown in Table~\ref{tab:enrollment}, when setting the training proportion as 10\%, among 10,241 authentic samples from 20 users, the average number of audio samples provided by each user during the enrollment is only 51. Since the average time length of the voice command is smaller than 3 seconds, the enrollment can be done in less than 3 minutes. Compared with the time overhead on deploying an Alexa skill which is up to 5 minutes~\cite{amazon:overhead}, requiring 3 minutes for enrollment is acceptable in real-world scenarios. 

\begin{table}[t]
\caption{Enrollment times per user.}
\label{tab:enrollment}
\begin{tabular}{p{1.5cm}<{\centering}|p{1.34cm}<{\centering}|p{1.3cm}<{\centering}|p{1.2cm}<{\centering}|p{0.8cm}<{\centering}}
\hline
Training proportion & Authentic samples & Time (mm:ss) & Accuracy (\%) & EER (\%)   \\ \hline
10\%    & 51    & 02:33  & 99.14  & 0.96 \\ \hline
20\%    & 103   & 05:09  & 99.47  & 0.55 \\ \hline
30\%    & 155   & 07:45  & 99.63  & 0.43 \\ \hline
50\%    & 263   & 13:09  & 99.84  & 0.17 \\ \hline
\end{tabular}
\end{table}

\subsection{Distinguish between Different Users}

Since \frameName is designed for liveness detection, we mainly consider the voice command generated by the electrical loudspeaker as a spoofing sample in this study. This subsection explores the feasibility of user classification.

\begin{figure}[t]
\centering
\includegraphics[width=0.65\linewidth]{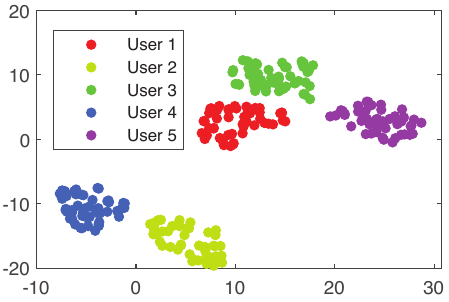}
\caption{Feature separation of 5 different users.}
\vspace{-4mm}
\label{fig:user:classification}
\end{figure}

We randomly select 250 authentic samples from 5 different users and then utilize t-Distributed Stochastic Neighbor Embedding (t-SNE) to reduce the dimension of their corresponding features.
As shown in Figure~\ref{fig:user:classification}, the feature vectors from different users are visually clustered after dimension reduction, which shows the feasibility of user classification. For all 10,241 authentic samples from 20 users, by leveraging two-fold cross-validation, \frameName achieves an overall speaker recognition accuracy of 99.88\%. Besides, the accuracy among different users ranges from 98.5\% to 100\%, which validates the effectiveness of \frameName on user authentication.

\subsection{Limitations and Countermeasures}
\label{sec:limitations}

We discuss some limitations of \frameName in this subsection.

\noindent \textbf{The user's burden on the enrollment.} We can incorporate the enrollment into daily use to reduce the user's time overhead on training \frameName. Firstly, the evaluation results from Section~\ref{sec:eval:conditions} show that \frameName is robust to the change of user's position, direction, and movement. That means the user can participate in the enrollment anytime. Then, to achieve this goal, we divide \frameName into working and idle phases. In the working phase,  when a user generates a voice command, \frameName collects the audio and saves the extracted features. During the idle phase, \frameName can automatically update the classifier based on these new generated features. These steps can be done automatically without human involvement, which means \frameName can continuously improve its performance along with daily use. 
However, we admit allowing the automatically continuous retraining process may involve other potential risks. For instance, attackers can launch poisoning attacks to reduce the performance of speech recognition and speaker verification~\cite{9519395,9519472,236234}.

\noindent \textbf{Impact of noise and other speakers.}
During the user's enrollment, we assume the environment is silent and there is no user who is talking. As shown in Section~\ref{sec:eval:noise}, since \frameName is a passive liveness detection that only depends on audios, the strong noise or other speaker's voice existing in the collected audios will inevitably degrade its performance.
Therefore, the existence of noise and other users who are talking will increase the enrollment time.
Fortunately, since \frameName is designed for the smart home or office environment, asking the users to keep a silent environment during enrollment is a reasonable assumption. We leave this issue as future work.

\noindent \textbf{Temporal stability of array fingerprint.}
 To evaluate the timeliness of \frameName, we recruit a participant to provide 100 authentic voice commands and launch voice spoofing per 24 hours. When using the classification model as described in Section~\ref{sec:eval:performance} and the audio dataset collected by 24 hours and 48 hours later, \frameName still achieves over 98\% accuracy. We admit that the generated feature may be variant when the participant changes her/his speaking manner or suffers from mood swings. As mentioned in Section~\ref{sec:limitations}, a feasible solution to address this issue is incorporating the enrollment into the user's daily use to ensure the freshness of the classification model of \frameName.

\section{Related Works}
\label{sec:related:work}

\noindent\textbf{Attacks on smart speakers.} The voice assistant is more vulnerable to the replay attack\cite{2017Monkey,Diao:2014:YVA:2666620.2666623,2014Jang,2015Petracca}. Apart from the classic replay attack, other advanced attacks are proposed. Firstly, the attacker can leverage medias including ultrasonic and laser to spoof voice assistance without incurring the user's perception \cite{dolphinAttack,2017song,2018Nirupam,2020Light}. Secondly, the subtle noises can be employed to generate the adversarial examples attacks~\cite{kreuk2018fooling,commandsong,2016Nicholas,2019kwon,2019zhou,2015Tavish,2018Lea}. Thirdly, several attacking methods can activate the malicious app to threaten the security of our smart home system\cite{gong2017crafting,2018Kumar,2019Danger,2019Life}.  Finally, Wang \etal~\cite{wangccs2020}propose modulated attack, which is the latest advanced voice spoofing method, and we evaluate it in Section~\ref{sec:eval:advanced}.

\noindent\textbf{Multi-factor based defenses.} 
As for the detecting method, some researches~\cite{vauth,2019Li,8433167} are based on wearable devices. Besides, several works utilize the Doppler effect ~\cite{2016AudioGest,2017Pittman}, gestures according to sound~\cite{2018VSkin}, or other biometry characteristics to deal with the security issue.
Lei \etal~\cite{8433167} and Meng \etal~\cite{Meng:2018:WES:3209582.3209591} proposed a wireless signal based method to thwart voice spoofing.
Lee \etal \cite{ubicomp} proposed a sonar-based solution to determine the user's AoA (angle of arrival) to do liveness detection. Zhang \etal \cite{zhang1,Zhang:2017:HYV:3133956.3133962} and Chen \etal \cite{magnetic} utilize the Doppler effect of ultrasonic and magnetic fields from loudspeakers as the essential characteristic for detecting attacks, respectively.
However, these methods either require the user to wear some specialized devices or utilize other devices (\eg, wireless sensors) to measure the environmental change caused by humans.

\noindent\textbf{Defenses relying on the collected audios.}
Shiota \etal \cite{shiota2015voice} and Wang \etal \cite{wang2019voicepop} utilized the Pop noise when the human speaks to differentiate the voice commands generated by real humans and devices. Yan \etal \cite{fieldprint2019} proposed the concept of using a fieldprint to detect spoofing attacks. Furthermore, Blue \etal \cite{blue2018hello} and Ahmed \etal \cite{void2020} utilized spectral power patterns to identify spoofing attacks alongside a single classification model to achieve lightness in authentication. Besides, in terms of feature selection, Defraene \etal \cite{2014Embedded} and Kamble \etal\cite{MFCC2018} propose novel spectrum-based features respectively. 
We analyze these passive liveness detection schemes in Section~\ref{sec:array:rq1}. 
Recently, Zhang \etal~\cite{eararray} propose EarArray to defend against ultrasonic-based attacks (\eg, dolphin attacks~\cite{dolphinAttack}), but it is not designed to detect spoofing audios with human voice frequency.

\section{Conclusion}
\label{sec:conclusion}

In this study, we propose a novel liveness detection system \frameName for thwarting voice spoofing attacks without any extra devices. We give a theoretical analysis of existing popular passive liveness detection schemes and propose a robust liveness feature \textit{array fingerprint}.
This novel feature both enhances effectiveness and broadens the application scenarios of passive liveness detection. 
\frameName is tested on both our MALD dataset and another public dataset, and the experimental results demonstrate \frameName is superior to existing passive liveness detection schemes. Besides, we evaluate multiple factors and demonstrate the robustness of \frameName.

\section*{Acknowledgments}
We thank the shepherd, Rahul Chatterjee, and other anonymous reviewers for their insightful comments. The authors affiliated with Shanghai Jiao Tong University were, in part, supported by the National Natural Science Foundation of China under Grant 62132013, and 61972453. Yuan Tian is partially supported by NSF award \#1943100. Haojin Zhu is the corresponding author.

{\footnotesize
\bibliographystyle{plain}
\bibliography{reference}

\begin{thebibliography}{10}

\bibitem{9519472}
Hadi Abdullah, Muhammad~Sajidur Rahman, Washington Garcia, Kevin Warren,
  Anurag~Swarnim Yadav, Tom Shrimpton, and Patrick Traynor.
\newblock Hear "no evil", see "kenansville"*: Efficient and transferable
  black-box attacks on speech recognition and voice identification systems.
\newblock In {\em 2021 IEEE Symposium on Security and Privacy (SP)}, pages
  712--729, 2021.

\bibitem{9519395}
Hadi Abdullah, Kevin Warren, Vincent Bindschaedler, Nicolas Papernot, and
  Patrick Traynor.
\newblock Sok: The faults in our asrs: An overview of attacks against automatic
  speech recognition and speaker identification systems.
\newblock In {\em 2021 IEEE Symposium on Security and Privacy (SP)}, pages
  730--747, 2021.

\bibitem{void2020}
Muhammad~Ejaz Ahmed, Il-Youp Kwak, Jun~Ho Huh, Iljoo Kim, Taekkyung Oh, and
  Hyoungshick Kim.
\newblock Void: A fast and light voice liveness detection system.
\newblock In {\em 29th {USENIX} Security Symposium ({USENIX} Security 20)},
  pages 2685--2702. {USENIX} Association, August 2020.

\bibitem{2017Monkey}
Efthimios Alepis and Constantinos Patsakis.
\newblock Monkey says, monkey does: Security and privacy on voice assistants.
\newblock {\em IEEE Access}, 5:17841--17851, 2017.

\bibitem{blue20182ma}
Logan Blue, Hadi Abdullah, Luis Vargas, and Patrick Traynor.
\newblock 2ma: Verifying voice commands via two microphone authentication.
\newblock In {\em Proceedings of the 2018 on Asia Conference on Computer and
  Communications Security}, page 89–100. ACM, 2018.

\bibitem{blue2018hello}
Logan Blue, Luis Vargas, and Patrick Traynor.
\newblock Hello, is it me you're looking for? differentiating between human and
  electronic speakers for voice interface security.
\newblock In {\em Proceedings of the 11th ACM Conference on Security \& Privacy
  in Wireless and Mobile Networks}, page 123–133. ACM, 2018.

\bibitem{2016Nicholas}
Nicholas Carlini, Pratyush Mishra, Tavish Vaidya, Yuankai Zhang, Micah Sherr,
  Clay Shields, David Wagner, and Wenchao Zhou.
\newblock Hidden voice commands.
\newblock In {\em 25th {USENIX} Security Symposium ({USENIX} Security 16)},
  pages 513--530, Austin, TX, August 2016. {USENIX} Association.

\bibitem{who:is:real:bob}
G.~Chen, S.~Chen, L.~Fan, X.~Du, Z.~Zhao, F.~Song, and Y.~Liu.
\newblock Who is real bob? adversarial attacks on speaker recognition systems.
\newblock In {\em 2021 IEEE Symposium on Security and Privacy (SP)}, pages
  55--72. IEEE Computer Society, may 2021.

\bibitem{magnetic}
S.~Chen, K.~Ren, S.~Piao, C.~Wang, Q.~Wang, J.~Weng, L.~Su, and A.~Mohaisen.
\newblock You can hear but you cannot steal: Defending against voice
  impersonation attacks on smartphones.
\newblock In {\em 2017 IEEE 37th International Conference on Distributed
  Computing Systems (ICDCS)}, pages 183--195, June 2017.

\bibitem{2014Embedded}
Bruno Defraene, Toon van Waterschoot, Moritz Diehl, and Marc Moonen.
\newblock Embedded-optimization-based loudspeaker compensation using a generic
  hammerstein loudspeaker model.
\newblock In {\em 21st European Signal Processing Conference (EUSIPCO 2013)},
  pages 1--5, 2013.

\bibitem{236234}
Ambra Demontis, Marco Melis, Maura Pintor, Matthew Jagielski, Battista Biggio,
  Alina Oprea, Cristina Nita-Rotaru, and Fabio Roli.
\newblock Why do adversarial attacks transfer? explaining transferability of
  evasion and poisoning attacks.
\newblock In {\em 28th {USENIX} Security Symposium ({USENIX} Security 19)},
  pages 321--338, Santa Clara, CA, August 2019. {USENIX} Association.

\bibitem{Diao:2014:YVA:2666620.2666623}
Wenrui Diao, Xiangyu Liu, Zhe Zhou, and Kehuan Zhang.
\newblock Your voice assistant is mine: How to abuse speakers to steal
  information and control your phone.
\newblock In {\em Proceedings of the 4th ACM Workshop on Security and Privacy
  in Smartphones \& Mobile Devices (SPSM)}, pages 63--74, 2014.

\bibitem{DBLP:conf/ccs/DingH18}
Wenbo Ding and Hongxin Hu.
\newblock On the safety of iot device physical interaction control.
\newblock In {\em Proceedings of the 2018 ACM SIGSAC Conference on Computer and
  Communications Security}, page 832–846, 2018.

\bibitem{9193961}
Yudi Dong and Yu-Dong Yao.
\newblock Secure mmwave-radar-based speaker verification for iot smart home.
\newblock {\em IEEE Internet of Things Journal}, 8(5):3500--3511, 2021.

\bibitem{vauth}
Huan Feng, Kassem Fawaz, and Kang~G. Shin.
\newblock Continuous authentication for voice assistants.
\newblock In {\em Proceedings of the 23rd Annual International Conference on
  Mobile Computing and Networking}, page 343–355. Association for Computing
  Machinery, 2017.

\bibitem{amazon:workflow}
Alexandre Gonfalonieri.
\newblock How amazon alexa works? your guide to natural language processing
  (ai).
\newblock
  \url{https://towardsdatascience.com/how-amazon-alexa-works-your-guide-to-natural-language-processing-ai-7506004709d3},
  2018.

\bibitem{gong2017crafting}
Yuan Gong and Christian Poellabauer.
\newblock Crafting adversarial examples for speech paralinguistics
  applications.
\newblock {\em arXiv preprint arXiv:1711.03280}, 2017.

\bibitem{remasc}
Yuan Gong, Jian Yang, Jacob Huber, Mitchell MacKnight, and Christian
  Poellabauer.
\newblock {ReMASC: Realistic Replay Attack Corpus for Voice Controlled
  Systems}.
\newblock In {\em Proc. Interspeech 2019}, pages 2355--2359, 2019.

\bibitem{hansen2001fundamentals}
Colin~H Hansen.
\newblock Fundamentals of acoustics.
\newblock {\em Occupational Exposure to Noise: Evaluation, Prevention and
  Control. World Health Organization}, pages 23--52, 2001.

\bibitem{amazon:overhead}
Amazon Inc.
\newblock Create and manage alexa-hosted skills.
\newblock
  \url{https://developer.amazon.com/en-US/docs/alexa/hosted-skills/alexa-hosted-skills-create.html},
  2021.

\bibitem{2014Jang}
Yeongjin Jang, Chengyu Song, Simon~P. Chung, Tielei Wang, and Wenke Lee.
\newblock A11y attacks: Exploiting accessibility in operating systems.
\newblock In {\em Proceedings of the 2014 ACM SIGSAC Conference on Computer and
  Communications Security}, CCS '14, page 103–115. Association for Computing
  Machinery, 2014.

\bibitem{MFCC2018}
M.~R. {Kamble} and H.~A. {Patil}.
\newblock Novel amplitude weighted frequency modulation features for replay
  spoof detection.
\newblock In {\em 2018 11th International Symposium on Chinese Spoken Language
  Processing (ISCSLP)}, pages 185--189, 2018.

\bibitem{Kinnunen2017}
Tomi Kinnunen, Md. Sahidullah, Héctor Delgado, Massimiliano Todisco, Nicholas
  Evans, Junichi Yamagishi, and Kong~Aik Lee.
\newblock The asvspoof 2017 challenge: Assessing the limits of replay spoofing
  attack detection.
\newblock In {\em Proc. Interspeech 2017}, pages 2--6, 2017.

\bibitem{kreuk2018fooling}
Felix Kreuk, Yossi Adi, Moustapha Cisse, and Joseph Keshet.
\newblock Fooling end-to-end speaker verification with adversarial examples.
\newblock In {\em 2018 IEEE International Conference on Acoustics, Speech and
  Signal Processing (ICASSP)}, pages 1962--1966. IEEE, 2018.

\bibitem{2018Kumar}
Deepak Kumar, Riccardo Paccagnella, Paul Murley, Eric Hennenfent, Joshua Mason,
  Adam Bates, and Michael Bailey.
\newblock Skill squatting attacks on amazon alexa.
\newblock In {\em Proceedings of the 27th USENIX Conference on Security
  Symposium}, SEC'18, page 33–47. USENIX Association, 2018.

\bibitem{2019kwon}
Hyun Kwon, Hyunsoo Yoon, and Ki-Woong Park.
\newblock Poster: Detecting audio adversarial example through audio
  modification.
\newblock In {\em Proceedings of the 2019 ACM SIGSAC Conference on Computer and
  Communications Security}, CCS '19, page 2521–2523. Association for
  Computing Machinery, 2019.

\bibitem{ubicomp}
Yeonjoon Lee, Yue Zhao, Jiutian Zeng, Kwangwuk Lee, Nan Zhang, Faysal~Hossain
  Shezan, Yuan Tian, Kai Chen, and XiaoFeng Wang.
\newblock Using sonar for liveness detection to protect smart speakers against
  remote attackers.
\newblock {\em Proc. ACM Interact. Mob. Wearable Ubiquitous Technol.}, 4(1),
  March 2020.

\bibitem{8433167}
X.~{Lei}, G.~{Tu}, A.~X. {Liu}, C.~{Li}, and T.~{Xie}.
\newblock The insecurity of home digital voice assistants - vulnerabilities,
  attacks and countermeasures.
\newblock In {\em 2018 IEEE Conference on Communications and Network Security
  (CNS)}, pages 1--9, May 2018.

\bibitem{2019Li}
Xiaopeng Li, Fengyao Yan, Fei Zuo, Qiang Zeng, and Lannan Luo.
\newblock Touch well before use: Intuitive and secure authentication for iot
  devices.
\newblock In {\em The 25th Annual International Conference on Mobile Computing
  and Networking}, MobiCom '19. Association for Computing Machinery, 2019.

\bibitem{alexa:review}
Divyang Makwana.
\newblock Amazon echo smart speaker (3rd gen) review.
\newblock
  \url{https://www.mobigyaan.com/amazon-echo-smart-speaker-3rd-gen-review},
  2020.

\bibitem{matrix}
Matirx.
\newblock Matrix creator.
\newblock
  \url{https://matrix-io.github.io/matrix-documentation/matrix-creator/overview/},
  2020.

\bibitem{Meng:2018:WES:3209582.3209591}
Yan Meng, Zichang Wang, Wei Zhang, Peilin Wu, Haojin Zhu, Xiaohui Liang, and
  Yao Liu.
\newblock Wivo: Enhancing the security of voice control system via wireless
  signal in iot environment.
\newblock In {\em Proceedings of the Eighteenth ACM International Symposium on
  Mobile Ad Hoc Networking and Computing (MobiHoc)}, pages 81--90, 2018.

\bibitem{2015Petracca}
Giuseppe Petracca, Yuqiong Sun, Trent Jaeger, and Ahmad Atamli.
\newblock Audroid: Preventing attacks on audio channels in mobile devices.
\newblock In {\em Proceedings of the 31st Annual Computer Security Applications
  Conference}, ACSAC 2015, page 181–190. Association for Computing Machinery,
  2015.

\bibitem{2017Pittman}
Corey~R. Pittman and Joseph~J. LaViola.
\newblock Multiwave: Complex hand gesture recognition using the doppler effect.
\newblock In {\em Proceedings of the 43rd Graphics Interface Conference}, GI
  '17, page 97–106. Canadian Human-Computer Communications Society, 2017.

\bibitem{BackDoor}
Nirupam Roy, Haitham Hassanieh, and Romit Roy~Choudhury.
\newblock Backdoor: Making microphones hear inaudible sounds.
\newblock In {\em Proceedings of the 15th ACM Annual International Conference
  on Mobile Systems, Applications, and Services (MobiSys)}, pages 2--14, 2017.

\bibitem{2018Nirupam}
Nirupam Roy, Sheng Shen, Haitham Hassanieh, and Romit~Roy Choudhury.
\newblock Inaudible voice commands: The long-range attack and defense.
\newblock In {\em 15th {USENIX} Symposium on Networked Systems Design and
  Implementation ({NSDI} 18)}, pages 547--560. {USENIX} Association, April
  2018.

\bibitem{2016AudioGest}
Wenjie Ruan, Quan~Z. Sheng, Lei Yang, Tao Gu, Peipei Xu, and Longfei Shangguan.
\newblock Audiogest: Enabling fine-grained hand gesture detection by decoding
  echo signal.
\newblock In {\em Proceedings of the 2016 ACM International Joint Conference on
  Pervasive and Ubiquitous Computing}, UbiComp '16, page 474–485. Association
  for Computing Machinery, 2016.

\bibitem{2018Lea}
Lea Sch{\"{o}}nherr, Katharina Kohls, Steffen Zeiler, Thorsten Holz, and
  Dorothea Kolossa.
\newblock Adversarial attacks against automatic speech recognition systems via
  psychoacoustic hiding.
\newblock In {\em 26th Annual Network and Distributed System Security
  Symposium}, pages 1--15. The Internet Society, 2019.

\bibitem{direction}
Sheng Shen, Daguan Chen, Yu-Lin Wei, Zhijian Yang, and Romit~Roy Choudhury.
\newblock Voice localization using nearby wall reflections.
\newblock In {\em Proceedings of the 26th Annual International Conference on
  Mobile Computing and Networking}, MobiCom '20. Association for Computing
  Machinery, 2020.

\bibitem{shiota2015voice}
Sayaka Shiota, Fernando Villavicencio, Junichi Yamagishi, Nobutaka Ono, Isao
  Echizen, and Tomoko Matsui.
\newblock Voice liveness detection algorithms based on pop noise caused by
  human breath for automatic speaker verification.
\newblock In {\em Sixteenth annual conference of the international speech
  communication association}, 2015.

\bibitem{2017song}
Liwei Song and Prateek Mittal.
\newblock Poster: Inaudible voice commands.
\newblock In {\em Proceedings of the 2017 ACM SIGSAC Conference on Computer and
  Communications Security}, CCS '17, page 2583–2585. Association for
  Computing Machinery, 2017.

\bibitem{respeaker}
Seeed Studio.
\newblock Respeaker core v2.0. 2019.
\newblock \url{http://wiki.seeedstudio.com/ReSpeaker_Core_v2.0/}, 2020.

\bibitem{2020Light}
Takeshi Sugawara, Benjamin Cyr, Sara Rampazzi, Daniel Genkin, and Kevin Fu.
\newblock Light commands: Laser-based audio injection attacks on
  voice-controllable systems.
\newblock In {\em 29th {USENIX} Security Symposium ({USENIX} Security 20)},
  pages 2631--2648. {USENIX} Association, August 2020.

\bibitem{2018VSkin}
Ke~Sun, Ting Zhao, Wei Wang, and Lei Xie.
\newblock Vskin: Sensing touch gestures on surfaces of mobile devices using
  acoustic signals.
\newblock In {\em Proceedings of the 24th Annual International Conference on
  Mobile Computing and Networking}, MobiCom '18, page 591–605. Association
  for Computing Machinery, 2018.

\bibitem{google:review}
Maggie Tillman.
\newblock Google home max review: Cranking smart speaker audio to the max.
\newblock
  \url{https://www.pocket-lint.com/smart-home/reviews/google/143184-google-home-max-review-specs-price},
  2019.

\bibitem{2015Tavish}
Tavish Vaidya, Yuankai Zhang, Micah Sherr, and Clay Shields.
\newblock Cocaine noodles: Exploiting the gap between human and machine speech
  recognition.
\newblock In {\em 9th {USENIX} Workshop on Offensive Technologies ({WOOT} 15)},
  Washington, D.C., August 2015. {USENIX} Association.

\bibitem{wang2019voicepop}
Qian Wang, Xiu Lin, Man Zhou, Yanjiao Chen, Cong Wang, Qi~Li, and Xiangyang
  Luo.
\newblock Voicepop: A pop noise based anti-spoofing system for voice
  authentication on smartphones.
\newblock In {\em IEEE INFOCOM 2019-IEEE Conference on Computer
  Communications}, pages 2062--2070. IEEE, 2019.

\bibitem{wangccs2020}
Shu Wang, Jiahao Cao, Xu~He, Kun Sun, and Qi~Li.
\newblock When the differences in frequency domain are compensated:
  Understanding and defeating modulated replay attacks on automatic speech
  recognition.
\newblock In {\em Proceedings of the 2020 ACM SIGSAC Conference on Computer and
  Communications Security}, CCS '20, page 1103–1119. Association for
  Computing Machinery, 2020.

\bibitem{fieldprint2019}
Chen Yan, Yan Long, Xiaoyu Ji, and Wenyuan Xu.
\newblock The catcher in the field: A fieldprint based spoofing detection for
  text-independent speaker verification.
\newblock In {\em Proceedings of the 2019 ACM SIGSAC Conference on Computer and
  Communications Security}, CCS ’19, page 1215–1229. Association for
  Computing Machinery, 2019.

\bibitem{commandsong}
Xuejing Yuan, Yuxuan Chen, Yue Zhao, Yunhui Long, Xiaokang Liu, Kai Chen,
  Shengzhi Zhang, Heqing Huang, XiaoFeng Wang, and Carl~A. Gunter.
\newblock Commandersong: A systematic approach for practical adversarial voice
  recognition.
\newblock In {\em 27th {USENIX} Security Symposium ({USENIX} Security 18)},
  pages 49--64, Baltimore, MD, August 2018. {USENIX} Association.

\bibitem{eararray}
Guoming Zhang, Xiaoyu Ji, Xinfeng Li, Gang Qu, and Wenyuan Xu.
\newblock Eararray: Defending against dolphinattack via acoustic attenuation.
\newblock In {\em NDSS}, 2021.

\bibitem{dolphinAttack}
Guoming Zhang, Chen Yan, Xiaoyu Ji, Tianchen Zhang, Taimin Zhang, and Wenyuan
  Xu.
\newblock Dolphinattack: Inaudible voice commands.
\newblock In {\em Proceedings of the 2017 ACM SIGSAC Conference on Computer and
  Communications Security (CCS)}, pages 103--117, 2017.

\bibitem{vmask}
Lei Zhang, Yan Meng, Jiahao Yu, Chong Xiang, Brandon Falk, and Haojin Zhu.
\newblock Voiceprint mimicry attack towards speaker verification system in
  smart home.
\newblock In {\em IEEE INFOCOM 2020 - IEEE Conference on Computer
  Communications}, pages 377--386, 2020.

\bibitem{viblive}
Linghan Zhang, Sheng Tan, Zi~Wang, Yili Ren, Zhi Wang, and Jie Yang.
\newblock Viblive: {A} continuous liveness detection for secure voice user
  interface in iot environment.
\newblock In {\em {ACSAC} '20: Annual Computer Security Applications
  Conference}, pages 884--896. {ACM}, 2020.

\bibitem{Zhang:2017:HYV:3133956.3133962}
Linghan Zhang, Sheng Tan, and Jie Yang.
\newblock Hearing your voice is not enough: An articulatory gesture based
  liveness detection for voice authentication.
\newblock In {\em Proceedings of the 2017 ACM SIGSAC Conference on Computer and
  Communications Security (CCS)}, pages 57--71, 2017.

\bibitem{zhang1}
Linghan Zhang, Sheng Tan, Jie Yang, and Yingying Chen.
\newblock Voicelive: A phoneme localization based liveness detection for voice
  authentication on smartphones.
\newblock In {\em Proceedings of the 2016 ACM SIGSAC Conference on Computer and
  Communications Security}, CCS ’16, page 1080–1091. Association for
  Computing Machinery, 2016.

\bibitem{2019Danger}
Nan Zhang, Xianghang Mi, Xuan Feng, XiaoFeng Wang, Yuan Tian, and Feng Qian.
\newblock Dangerous skills: Understanding and mitigating security risks of
  voice-controlled third-party functions on virtual personal assistant systems.
\newblock In {\em 2019 IEEE Symposium on Security and Privacy (SP)}, pages
  1381--1396. IEEE, 2019.

\bibitem{2019Life}
Yangyong Zhang, Lei Xu, Abner Mendoza, Guangliang Yang, Phakpoom
  Chinprutthiwong, and Guofei Gu.
\newblock Life after speech recognition: Fuzzing semantic misinterpretation for
  voice assistant applications.
\newblock In {\em NDSS}, 2019.

\bibitem{2019zhou}
M.~{Zhou}, Z.~{Qin}, X.~{Lin}, S.~{Hu}, Q.~{Wang}, and K.~{Ren}.
\newblock Hidden voice commands: Attacks and defenses on the vcs of autonomous
  driving cars.
\newblock {\em IEEE Wireless Communications}, 26(5):128--133, 2019.

\end{thebibliography}
}

\appendix

\begin{table*}[t]
\caption{Loudspeaker used for generating spoofing attacks.}
\centering
\begin{tabular}{c|c|c|c|c}
\hline
\textbf{No.} & \textbf{Type} & \textbf{Manufacture} &
\textbf{Model}& \textbf{Size (L*W*H in cm)}\\ \hline
1 & Loudspeaker & Bose & SoundLink Mini & 5.6 x 18.0 x 5.1 \\ \hline
2 & Tablet & Apple & iPad 6  & 24.0 x 16.9 x 0.7  \\ \hline
3 & Tablet &  Apple & iPad 9 & 24.0 $\times$ 16.9 $\times$ 0.7 \\ \hline
4 & Loudspeaker &  GGMM & Ture 360 & 17.5 $\times$ 10.9 $\times$ 10.9 \\ \hline
5 & Smartphone & Apple & iPhone 8 Plus & 15.8 x 7.8 x 0.7 \\ \hline
6 & Smartphone & Apple & iPhone 8 & 13.8 $\times$ 6.7 $\times$ 0.7 \\ \hline
7 & Smartphone & Apple & iPhone 6s & 13.8 $\times$ 6.7 $\times$ 0.7 \\ \hline
8 & Smartphone & Xiaomi & MIX2 & 15.2 $\times$ 7.6 $\times$ 0.8 \\ \hline
9 & Loudspeaker & Amazon & Echo Dot (2nd Gen) & 8.4 $\times$ 3.2 $\times$ 8.4 \\ \hline
10 & Laptop & Apple & MacBook Pro (2017) & 30.4 $\times$ 21.2 $\times$ 1.5 \\ \hline
11 & Loudspeaker &  VicTsing & SoundHot & 12.7 x 12.2 x 5.6 \\ \hline
12 & Loudspeaker & Ultimate Ears & Megaboom & 8.3 x 8.3 x 22.6 \\ \hline
13 & Loudspeaker & Amazon & Echo Plus (1st Gen) & 23.4 x 8.4 x 8.4 \\ \hline
14 & Smartphone & Xiaomi & Mi 9 & 15.8 $\times$ 7.5 $\times$ 0.8 \\ \hline
\end{tabular}
\label{table:device}
\end{table*}

\begin{table*}
\caption{Detailed information of MALD dataset.}
\centering
\begin{tabular}{p{1cm}<{\centering}|p{1.8cm}<{\centering}|p{1.6cm}<{\centering}|p{2.6cm}<{\centering}|p{7.6cm}<{\centering}}
\hline
\textbf{User \#} & \textbf{\# Authentic Samples } & \textbf{\# Spoofing Samples } &
\textbf{Distance (cm)}& \textbf{Spoofing Devices} \\ \hline
1, 7 & 1200 & 3600 & 60,120,180 & SoundLink Mini, iPad 6, iPhone 8 Plus\\ \hline
2 & 600 &  1079 & 60,120,180 & Ture360, iPhone 6s\\ \hline
3 & 533 &  904 & 60, 120, 180 & Ture360, iPad9 \\ \hline
4\textasciitilde6, 8 & 2305 &  6415 & 60, 120, 180 & iPad9, Ture360, MIX2\\ \hline
9\textasciitilde12 & 3211 &  3198 & 60, 120,180, 240 & Echo Plus (1st Gen)\\ \hline
13\textasciitilde 18 & 1191 & 4577 & 180 & iPad9, Mi 9, Echo Plus (1st Gen) \\ \hline
19 & 591 &  1767 & 60,120,180 & iPhone 8, Echo Dot (2nd Gen), MacBook Pro (2017)\\ \hline
20 & 610 &  998 & 60, 120, 180 & SoundHot, Megaboom\\ \hline
\end{tabular}
\label{table:overall}
\end{table*}

\section{LPCC Generation Process}
\label{appendix:lpcc}

For audio signal $y_k(t)$ collected by microphone $M_k$, to calculate the LPCC with the order $p = 15$, we firstly calculate the Linear Prediction Coding (LPC) as $a$:
\begin{equation}
  a = LPC(y_k(t),p),
\end{equation}
where $p$ is the order of LPC, and the collected LPC can be represented as $a = [a_0, a_1, \dots, a_p]$.
Then, for the LPCC coefficient $c = [c_0, c_1, \cdots, c_p]$, we have $c_0 = ln(p)$, and for other elments could be calculated as:
\begin{equation}
  c_n = -a_i - \sum_{k=1}^{i}{(1-\frac{k}{i}) a_k c_{i-k}}.
\end{equation}

\begin{figure}[t]
\centering
\includegraphics[width=8cm]{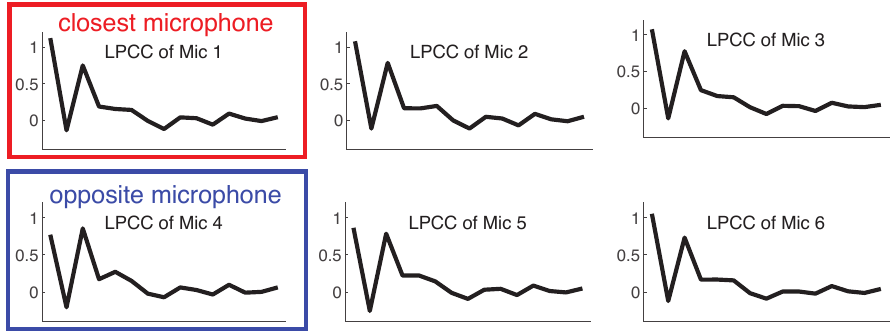}
\caption{LPCC in each channel.}
\label{fig:lpcc}
\end{figure}

In this study, the order $p$ is set to 15, 
and the LPCCs on each channel are shown in Figure~\ref{fig:lpcc}. In this figure, when $M_{1}$ is the closest microphone, for a microphone array with six channels, the opposite microphone is $M_4$. The LPCCs from these two channels are selected as $F_{LPC}$ in Section~\ref{sec:feature:lpc}.

\section{Dataset Descriptions}
\label{appendix:dataset}

\begin{figure}[t]
\centering
\includegraphics[width=0.5\linewidth]{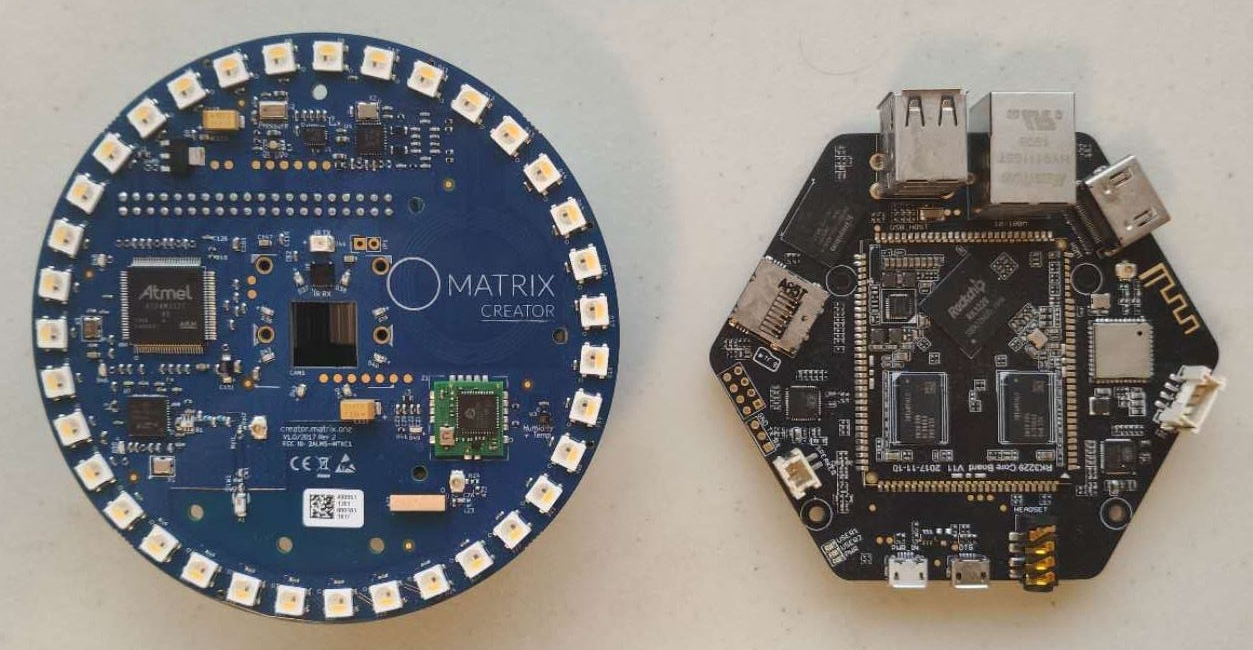}
\caption{Microphone array: Matrix Creator and Seeed ReSpeaker core V2.}
\label{fig:devices}
\end{figure}

First, the spoofing devices' information including manufacturing, model, and size is shown in Table~\ref{table:device}. Second, for each user, the data collection conditions including spoofing devices, distances, audio samples are summarized in Table~\ref{table:overall}. The dataset is collected by Matrix Creator and Seeed Respeaker core V2, which are shown in Figure~\ref{fig:devices}. Finally, we list the 20 voice commands used in our experiments as below:

    (1) OK Google.
    
    (2) Turn on Bluetooth.
    
    (3) Record a video.
    
    (4) Take a photo. 
    
    (5) Open music player.
    
    (6) Set an alarm for 6:30 am.
    
    (7) Remind me to buy coffee at 7 am.
    
    (8) What is my schedule for tomorrow?
    
    (9) Square root of 2105?
    
    (10) Open browser.
    
    (11) Decrease volume.
    
    (12) Turn on flashlight.
    
    (13) Set the volume to full.
    
    (14) Mute the volume.
    
    (15) What's the definition of transmit?
    
    (16) Call Pizza Hut.
    
    (17) Call the nearest computer shop.
    
    (18) Show me my messages.
    
    (19) Translate please give me directions to Chinese.
    
    (20) How do you say good night in Japanese?

\section{Experimental Details of Comparison with Existing Schemes}
\label{appendix:comparison}

\begin{table}[t]
\caption{Liveness detection performance under different classification models on the MALD dataset.}
\centering
\label{tab:models}
\begin{tabular}{c|c|c}
\hline
\multirow{2}{*}{Classifier type} & \multicolumn{2}{c}{Accuracy / EER (\%)}          \\ \cline{2-3} 
                                 & \frameName      & Mono feature\\ \hline
Neural network                   & 99.84 / 0.17 & 98.47 / 2.57                       \\ \hline
SVM-RBF                          & 99.48 / 1.07 & 98.81 / 1.78                       \\ \hline
kNN                              & 99.62 / 0.48 & 96.67 / 4.82                       \\ \hline
Decision tree                    & 96.35 / 5.97 & 94.84 / 7.34                       \\ \hline
\end{tabular}
\end{table}

When comparing \frameName with prior works, we strictly follow the steps described in \void~\cite{void2020} and \cafield~\cite{fieldprint2019}. 
In this section, we take \void as an example to show that \frameName is superior to existing schemes under various conditions.
More specifically, we add an experiment to explore the impact of different classifier models on the liveness detection performance of \frameName and \void.

We choose four different classification models: neural network, support vector machine with radial basis function kernel (SVM-RBF), k-Nearest Neighbor (kNN), decision tree. We fine-tune the parameters of each model. The results are shown in Table~\ref{tab:models}. It observed that \void achieves the best accuracy of 98.81\% when selecting SVM-RBF, which is the same as the paper~\cite{void2020}. These results prove \void is effective in detecting spoofing samples on \frameName dataset. However, it is observed that the performance of \frameName is better than that of \void under every classifier model. 
Besides, when applying these schemes on the third-party ReMasc Core dataset~\cite{remasc}, the performance of \frameName (\ie, the accuracy of 97.78\%)  is still better than that of \void (\ie, the accuracy of 84.37\%). In summary, compared with the mono channel-based scheme, exploiting multi-channel features achieves superior performance in the liveness detection task.

\section{Details of Modulated Attacks}
\label{appendix:inverse}

\begin{figure}[t]
\centering
\includegraphics[width=\linewidth]{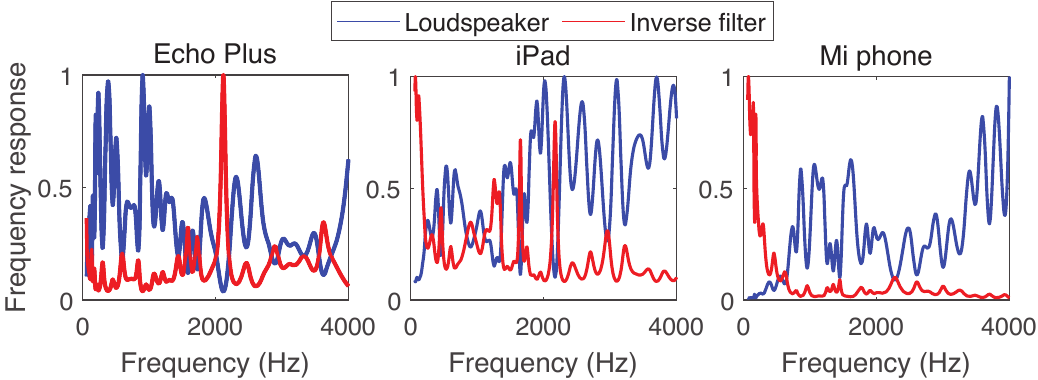}
\vspace{-4mm}
\caption{The amplitude responses of different spoofing devices and their corresponding inverse filters.}
\vspace{-4mm}
\label{fig:inverse}
\end{figure}

When re-implementing the modulated attack and calculating the detection accuracy of \frameName, we choose three spoofing devices \#3, \#13 and \#14 (\ie, iPad 9, Mi phone 9, and Amazon Echo Plus) as spoofing devices and Respeaker microphone array as the target device. To calculate the inverse filter for each device, we follow the steps described in the modulated attacks~\cite{wangccs2020}.  The frequency responses and their inverse filters of three spoofing devices are shown in Figure~\ref{fig:inverse}.

Then, after applying calculated inverse filters into the audios collected by the target device,  we generate 1,990, 1,791, and 1,994 modulated attack samples for Echo, iPad, and Mi respectively.

\end{document}